\documentclass[11pt,a4]{article}
\pdfoutput=1
\usepackage{jheppub}
\usepackage{graphicx,amssymb,amsmath,amsfonts}

\title{A rotating string model versus baryon spectra}

\author{Jacob Sonnenschein and Dorin Weissman}

\affiliation{The Raymond and Beverly Sackler School of Physics and Astronomy,\\
	Tel Aviv University, Ramat Aviv 69978, Israel}

\emailAdd{cobi@post.tau.ac.il}
\emailAdd{dorinw@mail.tau.ac.il}

\abstract{We continue our program of describing hadrons as rotating strings with massive endpoints. In this  paper we  propose models of baryons and confront them with the baryon Regge trajectories. We show that these are best fitted by a model of a single string with a quark at one endpoint and a diquark at the other. This model is preferred over the Y-shaped string model  with a quark at each endpoint.  We show how the model follows from a stringy model of the holographic baryon which  includes a baryonic vertex connected with $N_c$ strings to flavor probe  branes. From fitting to baryonic data we find that there is no clear evidence for a non-zero baryonic vertex mass, but if there is such a mass it should be located at one of the string endpoints. The available baryon trajectories in the angular momentum plane $(J,M^2)$, involving light, strange, and charmed baryons, are rather well fitted when adding masses to the string endpoints, with a single universal slope $\alpha\prime = $ 0.95 GeV$^{-2}$. Most of the results for the quark masses are then found to be consistent with the results extracted from the meson spectra in \cite{Sonnenschein:2014jwa}, where the value of the slope emerging from the meson fits  was found to be 0.90 GeV$^{-2}$. In the plane of radial excitations, $(n,M^2)$, we also find a good agreement between the meson and baryon slopes. The flavor structure of the diquark is examined, where our interest lies in particular on baryons composed of more than one quark heavier than the $u$ and $d$ quarks. For these baryons we present a method of checking the holographic interpretation of our results.}

\keywords{}
\def\be{\begin{equation}}
\def\ee{\end{equation}}
\newcommand{\del}{\partial}

\newcommand{\alp}{\ensuremath{\alpha'\:}}

\newcommand{\jp}[2]{\ensuremath{\frac{#1}{2}^{#2}}}
\newcommand{\jph}[2]{\ensuremath{#1/2^{#2}}}

\newcommand{\rchi}[1]{\ensuremath{\chi^2_m/\chi^2_l = #1}}
\newcommand{\ten}[1]{\times10^{#1}}

\begin{document}

\maketitle

\flushbottom

\section{Introduction}
The old idea that hadrons admit a stringy behavior has been reincarnated in recent years in the context of the holographic duality between gauge and string theories. Although the main focus of holography has been the description of properties of operators of the boundary gauge theory in terms of fields that reside either in the bulk or on flavor branes, strings in the bulk and strings ending on flavor branes have also played a major role in dualizing the gauge dynamics. For instance strings and $D_1$ branes with fixed endpoints on the boundary where shown to be the duals of  Wilson and 't Hooft lines respectively. In \cite{Sonnenschein:2014jwa} it was argued that the basic property of the  mesonic spectra, the Regge trajectories, cannot generically be accounted for by fields (in this case the fluctuations of the flavor branes) whereas holographic rotating strings naturally admit this behavior. Using a map developed in \cite{Kruczenski:2004me} we transformed the holographic strings of mesons into rotating strings with massive endpoints in flat space-time. This model, which is characterized by the string tension, the endpoint masses, and an intercept, was compared to PDG data and was found to be a universal setup describing mesons built from light quarks as well as those constructed from heavy ones.

It is very well known that mesons in nature admit Regge trajectories, but it is a lesser known fact that baryons also furnish Regge trajectories\cite{Inopin:1999nf}\cite{Tang:2000tb}\cite{Klempt:2009pi}. Thus, the spectra of baryons that follow from holography should admit such behavior. Like for the mesons, there are two candidates  for holographic duals of the baryons. Namely, they are either a field configuration or a string configuration. It was shown \cite{Hata:2007mb} that in terms of the former, baryons correspond to flavor instantons which are a static configuration of the 5 dimensional flavor gauge theory that resides on the $N_f$ flavor branes, compactified, for instance in the Sakai-Sugimoto model \cite{SakSug}, on an $S^4$. It was worked out, mainly in the context of a generalization of that model, that these instantons can adequately describe the static properties of baryons, and be responsible for appropriate (large $N_c$) nuclear interactions and nuclear matter \cite{Kaplunovsky:2010eh}\cite{Kaplunovsky:2012gb}\cite{Kaplunovsky:2013iza}. However, the spectrum of these instantons does not admit the Regge trajectories behavior neither for $M^2$ as a function of the angular momentum  $J$, nor as a function of the radial excitation number $n$. Therefore, we are led again to consider strings as the  holographic duals of baryons.

Since a single string that ends on the boundary or on a flavor brane corresponds to an external or a dynamical quark respectively, a stringy baryonic configuration has to connect $N_c$ strings. It was shown in \cite{Witten:1998zw} for the $AdS_5\times S^5$ background  and later for confining backgrounds \cite{Brandhuber:1998xy}\cite{Callan:1999zf} that a $D_p$ brane that wraps a non-trivial $p$ cycle with an RR flux of $N_c$ flowing out of it, must be connected to $N_c$ strings. The other end of each of these strings can be either on the boundary or on the flavor brane, thus  constituting an external or dynamical baryon respectively. Whereas in the original model \cite{Witten:1998zw} the baryonic vertex is a $D_5$ wrapping the fluxed $S^5$, in the confining models of \cite{SakSug} it is a $D_4$ brane wrapping an $S^4$, and in the model of \cite{Dymarsky:2009cm} it is a $D_3$ wrapping the three cycle of the deformed conifold. The location of the baryonic vertex in the radial direction and on the flavor brane world-volume coordinates is determined by minimizing the energy of the configuration \cite{Seki:2008mu}. In section \ref{sec:holobaryon} we briefly review two possible setups of stringy baryons: the spherically symmetric one and the totally asymmetric configuration. 

As we did in \cite{Sonnenschein:2014jwa} for the mesons while following \cite{Kruczenski:2004me}, we map the baryonic holographic configurations to configurations of strings with a baryonic vertex and massive endpoints in flat space-time. In this paper, like in \cite{Sonnenschein:2014jwa}, we take these stringy configurations in flat space-time as our theoretical models to compare with experimental data. We would like to emphasize that we do not provide a systematic controlled transformation from the holographic domain characterized by large $N_c$, large $\lambda$, and $\frac{N_f}{N_c}<<1$  to the realistic regime of $N_c=3$, $\lambda$ of order one, and $N_f=6$.
 The comparison between the predictions of our model and the experimental data enables us to decide what is the preferable configuration (in terms of the location of the baryonic vertex) and furthermore to extract the parameters that characterize the models. These parameters are the string tension $T$ (or equivalently its inverse, the slope of the trajectories \alp),  the endpoint masses $m_{sep}$, the mass of the baryonic vertex $m_{bv}$ and the intercept $a$. Whereas the former parameters show up  in the classical stringy model, the intercept emerges only upon quantization. It is well known that for a string with no massive endpoints, namely the case of the  linear Regge trajectory, the passage  from the  classical to quantum trajectories is via the replacement $J= \alp E^2 \rightarrow J+n= \alp E^2 + a$, where $n$ is the quantum radial excitation number. The relevant questions to our analysis are what is the theoretical value of the intercept for the massless case and moreover how does it depend on the string endpoint masses. In  \cite{Hellerman:2013kba} it was found, somewhat surprisingly, that for the case of a rotating bosonic string with angular momentum in a single plane, the intercept of the massless case is independent of the dimensionality of the space-time $D$ and takes the value of $\frac{D-2}{24}+ \frac{26-D}{24}= 1$. The first contribution to the intercept is the usual ``Casimir term" and the second one is the Polchinski-Strominger term.
  For the rotating string with massive endpoints a similar determination of the intercept has not yet been written down even though certain aspects of the quantization of such a system have been addressed \cite{Chodos:1973gt}\cite{Baker:2002km}\cite{Zahn:2013yma}.

%%%%%

The main result of the paper is that the best model providing a universal setup of the baryon is the model of a single string with a diquark at one endpoint and a quark at the other. This model is preferred over the model of a Y-shaped string with a quark at each endpoint. There is no clear evidence for a non-zero baryonic vertex mass, but if there is such a mass it should be located at one of the string endpoints. We see that we can fit the available baryon trajectories in the angular momentum plane \((J,M^2)\) rather well when adding masses to the endpoints, and we can do it, if we wish, with a single universal slope \(\alp = 0.95\) GeV\(^{-2}\). Most of the results for the quark masses are then found to be consistent with the results extracted from the meson spectra in \cite{Sonnenschein:2014jwa}, where the value of the slope emerging from the meson fits - 0.90 GeV\(^{-2}\) - is close to the value obtained here for the baryons. In the plane of string  excitations, \((n,M^2)\), we fitted the trajectories of light baryons and found that there too there is a good agreement between the meson and baryon slopes.

The paper is organized as follows. In section \ref{sec:holobaryon} we briefly review the concept of baryonic stringy configurations in holography.  Section \ref{sec:basic} is devoted to the basic (flat space-time) theoretical model for the baryonic string with massive endpoints. In section \ref{sec:models} we describe the various fitting  models, and then move on to the results of the fits in section \ref{sec:results}. Section \ref{sec:summary} begins with a summary of the results, compares them with the results of the meson fits, and offers a discussion of their implications regarding the structure and composition of the baryons. In our discussion of the structure we also look into (but do not analyze in detail) the decay modes of the baryons. In section \ref{sec:conclusions} we conclude, summarize and mention several future directions of the research program.

\section{Holographic stringy baryons} \label{sec:holobaryon}

\begin{figure}[t!] \centering
					\includegraphics[width=.75\textwidth, natheight=1029bp,natwidth=2000bp]{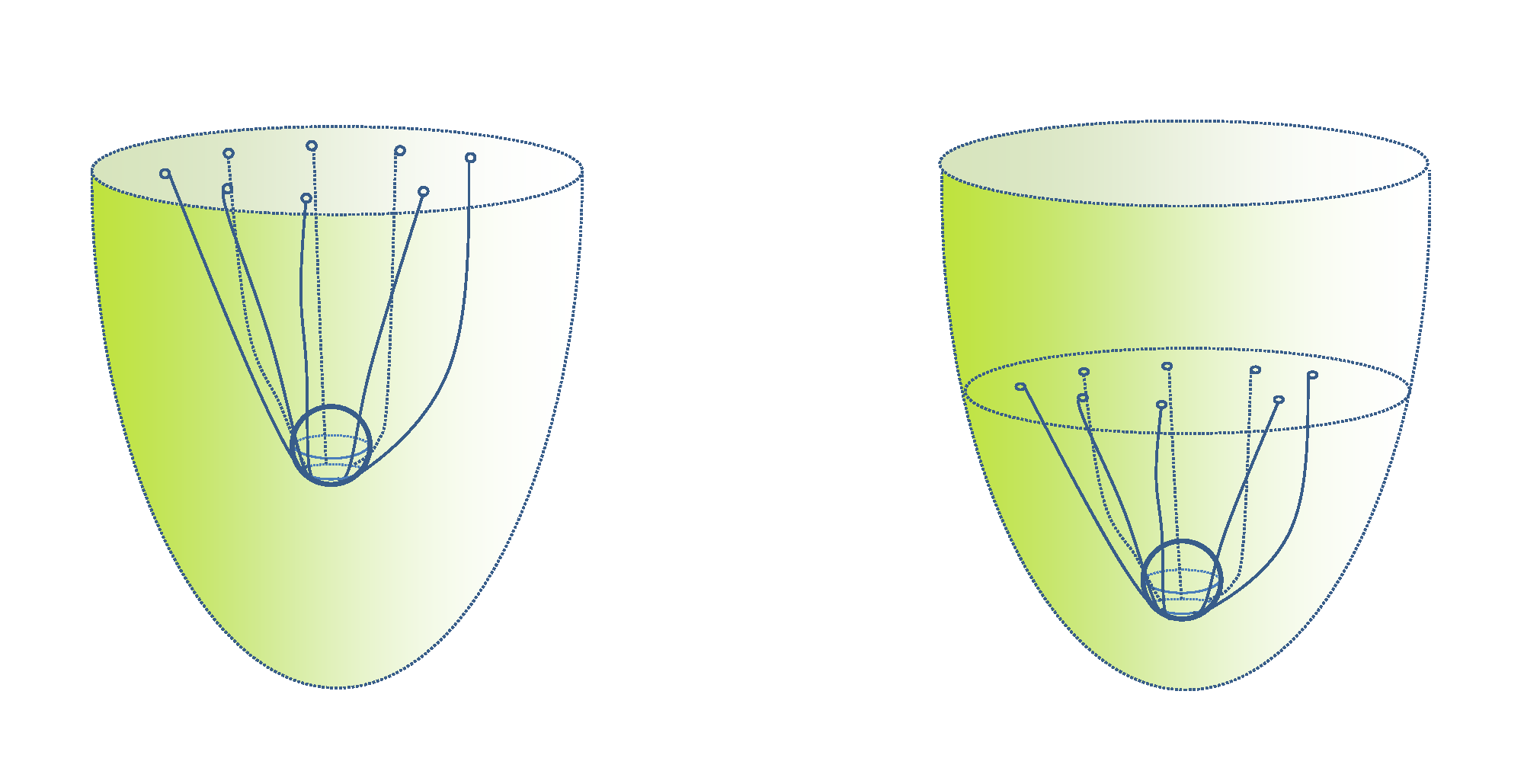}
					\caption{\label{schembaryons}  Schematic picture of holographic baryons. On the left is an external baryon with strings that end on the boundary, while on the right is a dynamical baryon with strings ending on a flavor brane.}
\end{figure}

A string that stretches from the boundary of a holographic background corresponds in the dual field theory to an external quark. Similarly a string that stretches  from  a flavor probe brane is a dual of a dynamical quark.  The dual of the  meson gauge singlet  thus corresponds to a string that starts and ends either on the boundary (external) or on a flavor brane (dynamical). It is thus clear that a stringy holographic baryon has to include $N_c$ strings that are connected together and end on a flavor brane  (dynamical baryon) or on the boundary (external baryon). The question is what provides the ``baryonic vertex'' that connects together $N_c$ strings.
In \cite{Witten:1998xy} it was shown  that in the $AdS_5\times S^5$ background, which is equipped with  an RR flux of value $N_c$,  a $D5$ brane that wraps the $S^5$ has to have  $N_c$ strings attach to it. This property can be generalized to other  holographic  backgrounds so that   a $D_p$ brane wrapping a non-trivial $p$ cycle with a  flux of an RR field  of value $N_c$  provides a baryonic vertex.   These two possible stringy configurations are
schematically depicted in figure (\ref{schembaryons}). Whereas the dual of  the original proposal \cite{Witten:1998xy} was a conformal field theory, Baryons can be constructed also in holographic backgrounds that correspond to confining field theories. A prototype model of this nature is the model of  $N_c$ $D4$ branes background  compactified on a circle with an $N_f$ $D8$--$5\overline{D8}$  U-shaped flavor branes \cite{SakSug}. In this model the baryonic vertex is made out of a $D4$ brane that wraps an $S^4$.  Another model for baryons in  a confining background is  the deformed conifold model  with $D7$--$\overline{D7}$ U-shaped flavor branes \cite{Dymarsky:2010ci}. In this model the baryon is a $D3$ brane that wraps the three-cycle of the deformed conifold.

The argument why a  $D_p$ brane wrapping a fluxed  $p$ cycle  is a baryonic vertex is in fact very simple.  The  world-volume action of the wrapped $D_p$ brane has the form of  $S= S_{DBI}+S_{CS}$. The CS term takes the following form
\be
S_{CS}= \int\limits_{S_p\times R_1} \!\! \sum_i c_{p_i} \wedge e^{F}
= \int\limits_{S_p\times R_1} \!\!\! c_{p-1}\wedge F
= -\!\!\!\! \int\limits_{S_p\times R_1} \!\!\! F_p\wedge A
= -N_c\int\limits_{R_1} \!\! A
\ee
where (i) the sum over \(i\) is over the RR $p$-forms that  reside in the background, (ii) from the sum one particular RR form  was chosen,  the $c_{p-1}$-form that couples to the Abelian field-strength F,  (iii) for simplicity's sake we took the $p$ cycle to be $S^p$, (iv) $A$  is the Abelian  connection  that resides on the wrapped brane, (v) $F_p$ is the RR $p$-form field strength, and (vi) in the last step we have made use of the fact that $\int_{S^p} {F_p} = N_c$.
This implies that there is a charge $N_c$  for the Abelian gauge field. Since in a compact space one cannot have non-balanced charges and since the endpoint of a string carries a charge one, there  must be $N_c$  strings attached to it. It is interesting to note that a baryonic vertex rather than being a ``fractional'' $D_0$ brane of the form of a $D_p$ brane wrapping a $p$ cycle, can also be a $D_0$ brane in an $N_c$ fluxed background. This is the case in the non-critical string backgrounds like \cite{Kuperstein:2004yf} where there is no non-trivial cycle to wrap branes over, but an ``ordinary" $D_0$ brane in this background will also have a CS term of the form $N_c\int_{R_1} A$.

\begin{figure}[t!] \centering
					\includegraphics[width=.80\textwidth, natheight = 549bp, natwidth = 831bp]{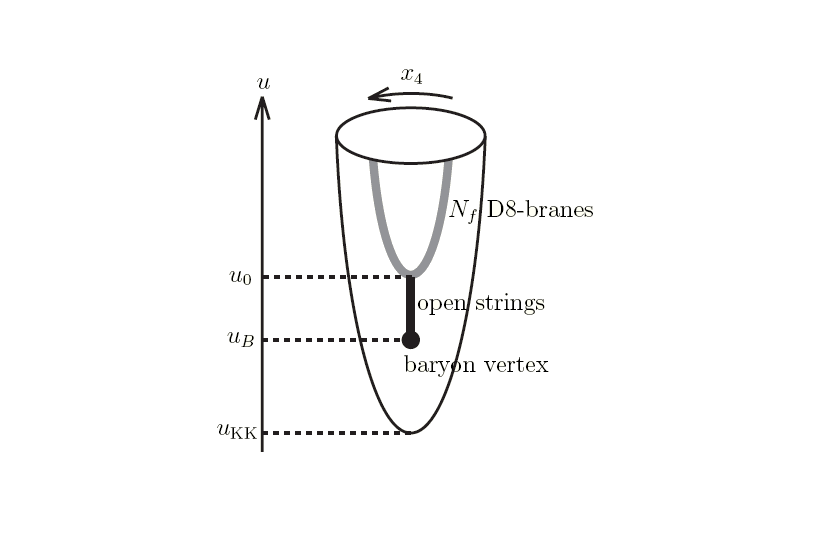}
					\caption{\label{baryonloc} The location of the baryonic vertex  along the radial direction of the Sakai-Sugimoto model. $u_0$ is the tip of the flavor branes and $u_{kk}$ is the ``confining scale" of the model.}
\end{figure}

Next we would like to determine the  location of the baryonic vertex in the radial dimension.  In particular the question is whether it is located on the flavor branes or below them.   This  is schematically depicted in figure (\ref{baryonloc}) for the model of \cite{SakSug}.

 In \cite{Seki:2008mu} it was shown by minimizing the ``mechanical energy" of the $N_c$ strings and the wrapped brane that it is preferable for the baryonic vertex to be located on the flavor branes in the model of \cite{SakSug}.  For the baryonic vertex of the model of \cite{Dymarsky:2009cm} it was shown that if the tip of the U-shaped flavor brane is close to the  lowest point of the deformed conifold the baryonic vertex does  dissolve in the flavor  branes. It is interesting to note that  for a background that corresponds to the  deconfining phase of the dual gauge theory the baryonic vertex falls into the ``black hole" and thus the baryon dissolves.

\begin{figure}[t!] \centering
					\includegraphics[width=.75\textwidth, natheight = 784bp, natwidth = 1374bp]{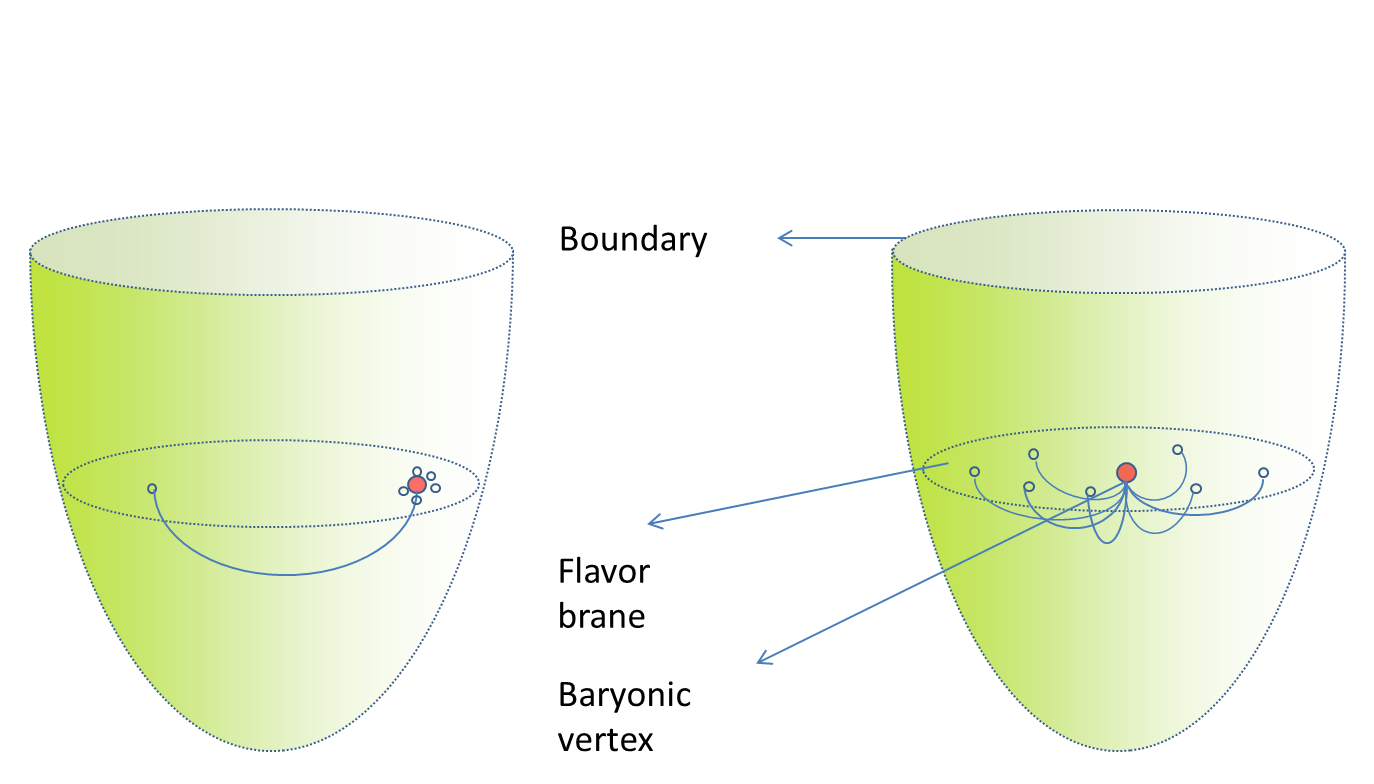}
					\caption{\label{fig:holbaryon} The location of the baryonic vertex on the flavor brane and the corresponding configuration of the baryon for some large \(N_c\). When \(N_c = 3\), the left is the analog of the quark-diquark flat space-time model, and the right the analog of the Y-shape model.}
\end{figure}

The locations of the baryonic vertex and the ends of the $N_c$ strings  on the flavor branes is  a dynamical issue. In figure (\ref{fig:holbaryon}) we draw two possible setups. In one the baryonic vertex is located at the center and the $N_c$ ends of the strings are located around it in a spherically symmetric way. In the ``old" stringy model of baryons for three colors this is the analog of the Y-shape configuration that we will further discuss in the sections describing the flat space-time models. Another possibility is that of a baryonic vertex connected by $N_c-1$  very short strings and with one long string  to the flavor branes. Since the product of  $N_c-1$  fundamental representations includes the anti-fundamental one, this configuration can be viewed as a string connecting a quark with an anti-quark. For the case of $N_c=3$ this is the analog of what will be discussed below as the quark-diquark stringy configuration. This latter string configuration (for any $N_c$) is similar to the stringy meson, but there is a crucial difference, which is the fact that the stringy baryon includes a baryonic vertex.

\begin{figure}[t!] \centering
					\includegraphics[width=.90\textwidth, natheight=783bp, natwidth=1926bp]{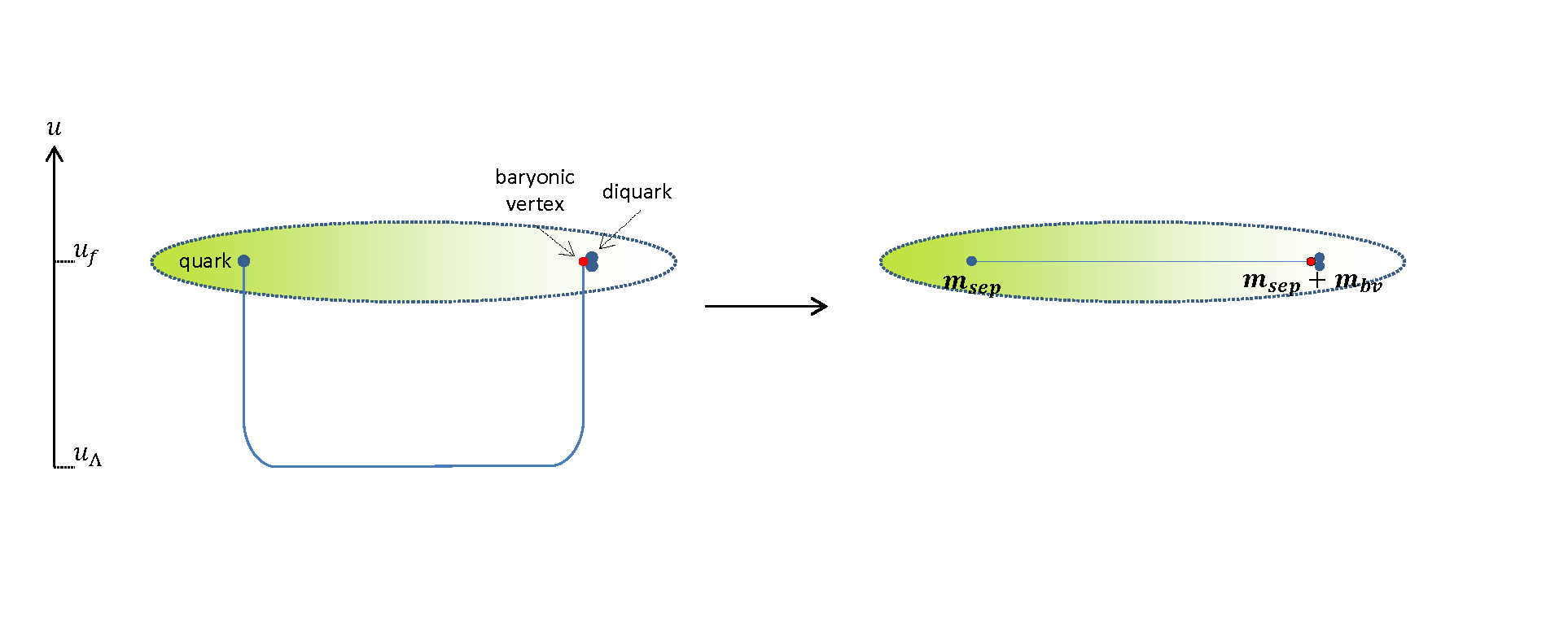}
					\caption{\label{holtoflat}  The holographic setup downlifted to $N_c=3$ of a quark and a diquark is mapped to a similar configuration in flat space-time. The vertical segments of the holographic string is mapped into masses of the endpoints.}
\end{figure}

It was shown in \cite{Kruczenski:2004me} that the classical rotating   holographic stringy configuration of the meson can be mapped into that of a classical rotating bosonic string in flat space-time with massive endpoints. A similar map applies also to  holographic stringy baryons that can be transformed into stringy baryons in flat space time with massive endpoints. We will proceed now to discuss this map for the central and asymmetric layouts of figure (\ref{fig:holbaryon}). The asymmetric holographic configuration of a quark and $N_c-1$ quarks on the two ends of the holographic string depicted in figure (\ref{holtoflat}) is mapped into a similar stringy configuration in flat space-time where the vertical segments of the string are transferred into massive endpoints of the string. On the left hand side of the string in flat space-time there is an endpoint with mass $m_{sep}$ given by
\be
m_{sep} = T\int_{u\Lambda}^{u_f} du \sqrt{g_{00}g_{uu}}
\ee
where $u_\lambda$ is the location of the ``wall'', $u_f$ is the location of the flavor brane, and $g_{00}$ and $g_{uu}$ are the $00$ and $uu$ components of the metric of the background.

On the right hand side the mass of the endpoint is $m_{sep}+m_{bv}$. This is the sum of the energy associated with the vertical segment of the string, just like that of the left hand side, and the mass of the baryonic vertex. Note that even though on the right endpoint of the string there are in fact two endpoints or ``two quarks" the mass is that of a single quark since there is a single vertical string segment, belonging to the string connecting the baryonic vertex with the lone quark at the other endpoint. This string setup is obviously very similar to that of the meson. The only difference is the baryonic vertex that resides at the diquark endpoint.  Since we do not know how to evaluate the mass of the baryonic vertex, it will be left as a free parameter to determine by the comparison with data. Our basic task in this case will be to distinguish between two options: (i) the mass of the baryonic vertex is much lighter than the endpoint mass, $m_{bv}<<m_{sep}$, in which case the masses at the two endpoints will be roughly the same, and (ii) an asymmetric setup with two different masses if the mass of the baryonic vertex cannot be neglected.

\begin{figure}[t!] \centering
					\includegraphics[width=.90\textwidth, natheight=882bp, natwidth=1845bp]{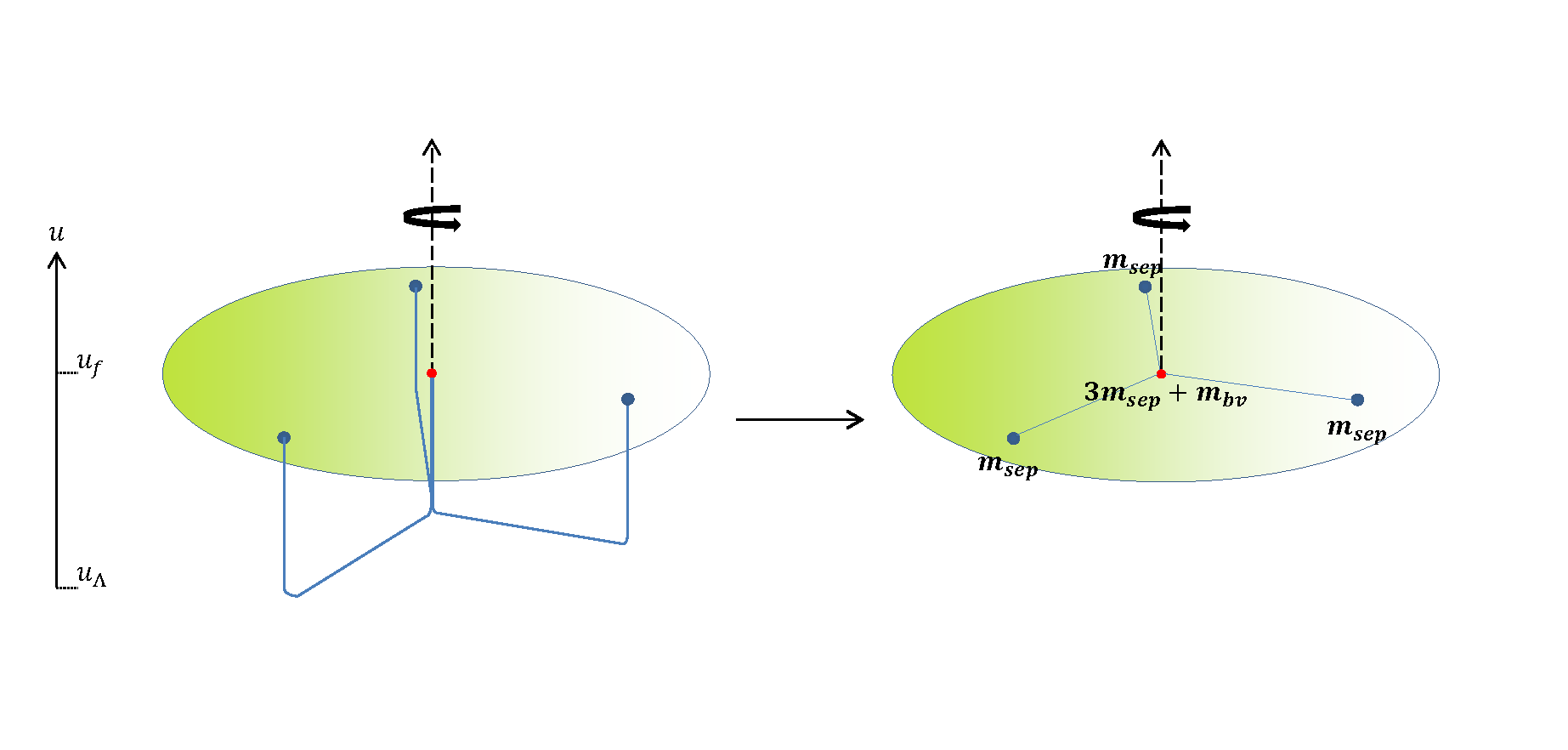}
					\caption{\label{centerholtoflat}   The map between the holographic central configuration (for $N_c=3$) and  the Y-shaped string in flat space-time.}
\end{figure}

The configuration with a central baryonic vertex  can be mapped into the analog of a Y-shaped object with $N_c$ massive endpoints and with a central baryonic vertex of mass $m_c= m_{bv}+N_c m_{sep}$. The factor of $N_c$ is due to the fact that there are $N_c$ strings that stretch from it vertically from the flavor brane to the ``wall", as can be seen in figure (\ref{centerholtoflat}) for the case of $N_c=3$. In this case, regardless of the ratio between the mass of the baryonic vertex and that of the string endpoint, there is a massive center which is at least as heavy as three sting endpoints.

\begin{figure}[ht!] \centering
					\includegraphics[width=.90\textwidth, natheight=1077bp,natwidth=2000bp]{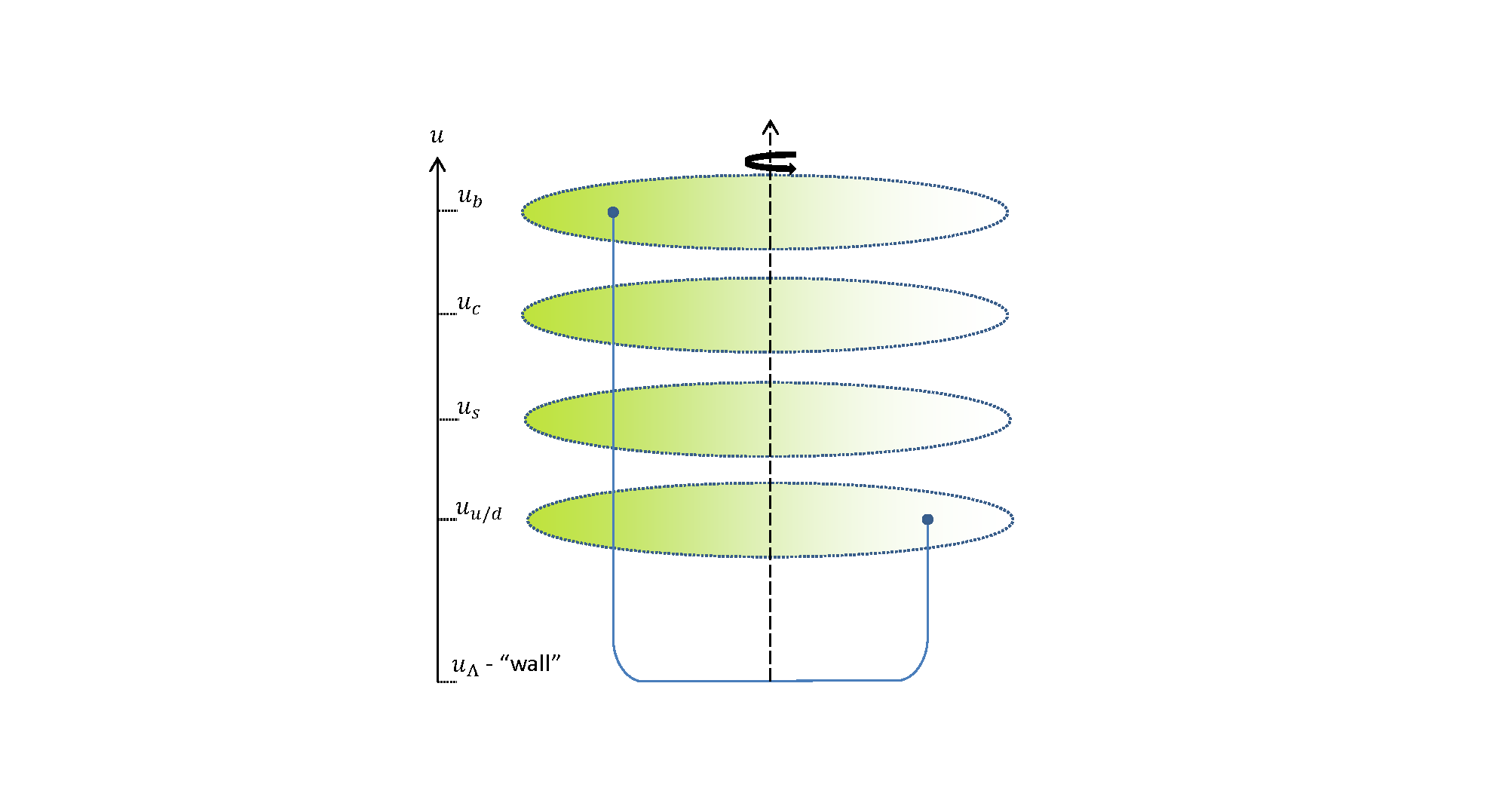}
					\caption{\label{udscbflavor}  Holographic setup with flavor branes associated with the $u,d$ quark, $s$, $c$ and $b$ quarks.}
\end{figure}

So far we have considered stringy holographic baryons that attach to one flavor brane. In holographic backgrounds one can introduce flavor branes at different radial locations thus corresponding to different string endpoint masses, or different quark masses. For instance, a setup that corresponds to $u$ and $d$ quarks of the same $m_{sep}$ mass, a strange $s$ quark, a charm $c$ quark, and a bottom $b$ quark is schematically drawn in figure (\ref{udscbflavor}).  A $B$ meson composed of a bottom quark and a light \(\bar{u}/\bar{d}\) anti-quark was added to the figure.

 Correspondingly there are many options of holographic stringy baryons that connect to different flavor branes. This obviously relates to baryons that are composed of quarks of different flavor. In fact there are typically more than one option for a given baryon. In addition to the distinction between the central and quark-diquark configurations  there are more than one option just to the latter configuration. We will demonstrate this situation in section \ref{sec:structure}, focusing on the case of the doubly strange $\Xi$ baryon (\(ssd\) or \(ssu\)). The difference between the the two holographic setups is translated to the two options of the diquark being either composed of two $s$ quarks, whereas the other setup features a $ds$ or $us$ diquark. Rather than trying to determine the preferred configuration from holography we will use a comparison with experimental data to investigate this issue.

\subsection{The stringy models in flat space-time and their stability} \label{sec:stability}

There are some words to be said about the models emerging when mapping the holographic rotating strings to flat space-time. In section \ref{sec:basic}, we will write the equations of motion of the string with massive endpoints in flat space-time and present the rotating solution. In this section, we briefly discuss another matter: the stability of the rotating solution. Specifically, we claim that the Y-shape model of the baryon is (classically) unstable. In our analysis of the spectrum we disregard this potential instability of the model and use the expressions for the energy and angular momentum of the unperturbed rotating solution of the Y-shape model as one of our fitting models, but it is important to remember that there is this theoretical argument against it as a universal setup for baryons before we test it out as a phenomenological model.

\begin{figure}[t!] \centering
					\includegraphics[width=.90\textwidth, natheight=592bp,natwidth=1500bp]{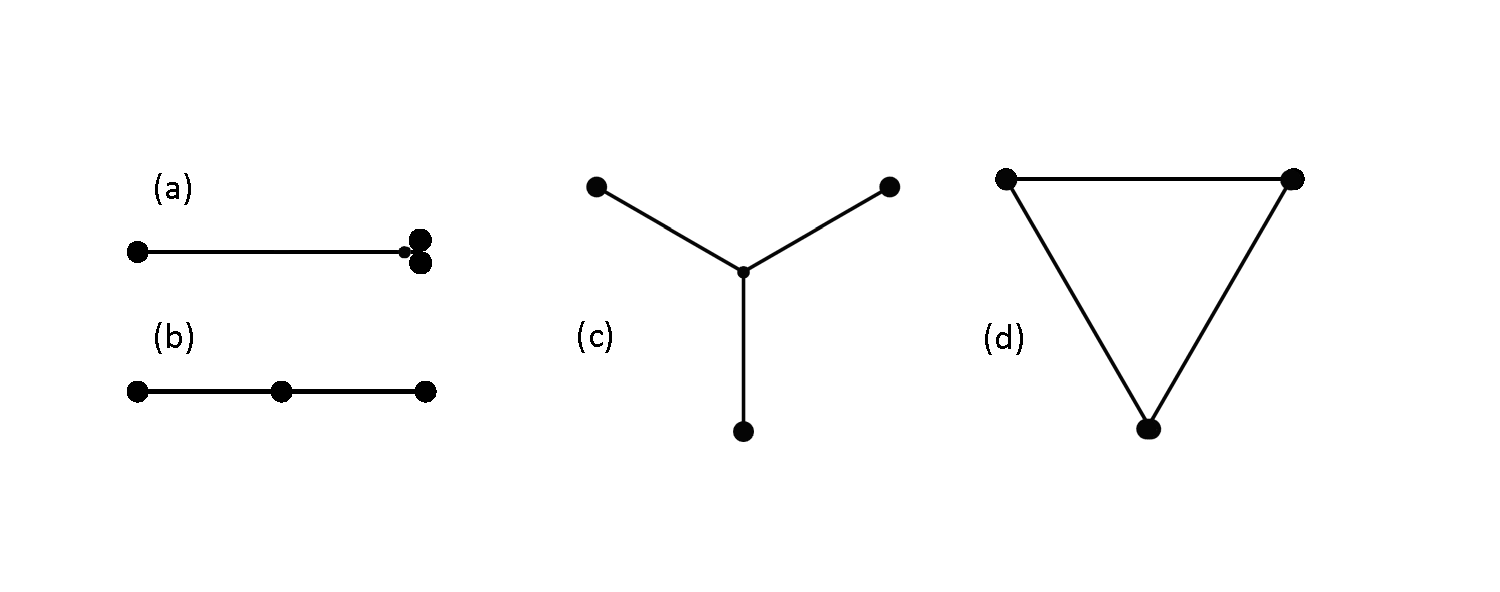}
					\caption{\label{fig:flatbaryons} Four different stringy models for the baryon in flat space-time. (a) is the quark-diquark model, (b) the two-string model, (c) the three-string Y-shape, and (d) the closed \(\Delta\)-shape. (a) and (c) are the models discussed in the preceding discussion of holographic strings and their mappings to flat space-time.}
\end{figure}

Other than the Y-shape and quark-diquark configurations which we analyze, there are two more possible stringy models for the baryon when considering a purely flat space-time point of view. These are drawn, together with the quark-diquark and Y-shape models, in figure (\ref{fig:flatbaryons}). The additional models are the two-string model where one of the quarks is located at the center of the baryon, and the other two are attached to it by a string. The second is the \(\Delta\)-shape model, in which each quark is connected to the other two. It can be looked at as a closed string with three points along it that carry finite momentum (which may be either massive or massless). While the two-string model may have its analog in holography, with the baryonic vertex lying with the quark at the center of mass, the \(\Delta\)-shaped string cannot be built if we impose the constraint that the three quarks should all be connected to a baryonic vertex.

Two independent analyses of the three-string model \cite{'tHooft:2004he}\cite{Sharov:2000pg} concluded that the rotating solution of the Y-shape configuration is unstable, even before taking quantum effects into account. In another work, it was found that the instability does not show itself in first order perturbation theory \cite{ThesisFederovsky}, but the claim of the unstable nature of the Y-shape model has been verified using numerical methods in \cite{ThesisHarpaz}, where the instability was observed in simulations and its dependence on endpoint masses was examined. To summarize the results, in the three-string Y-shape model, a perturbation to one of the three arms would cause it to shorten until eventually the Y-shape collapses to a form like that of the straight two-string model. From this model in turn a different kind of instability is expected \cite{'tHooft:2004he}\cite{Sharov:2000pg}. The quark in the center of the baryon will move away from the center of mass given a small perturbation and as it approaches one of the quarks at the endpoints quantum effects will induce a collapse to the single string quark-diquark model, it being energetically favorable for two of the quarks to form an diquark bound state in the anti-fundamental color representation. It would seem that all other models have an instability that would cause them to eventually collapse to the quark-diquark configuration as two quarks get close enough to each other.\footnote{More discussion and detailed analyses of the different stringy models of the baryon and their stability are found in the work of G.S. Sharov, most recently in \cite{Sharov:2013tga}.}

From a phenomenological point of view the models differ mainly in their prediction of the slope of the Regge trajectory. Assuming the strings in baryons have the same tension as those in mesons, we can see which of the models offers the best match. We will see that it is the configuration we know to be stable, that of a single string with a quark and a diquark at its endpoints. Therefore, our fitting analysis will be focused on this model.\footnote{The quark-diquark model was also used to analyze the baryon spectrum in \cite{Selem:2006nd}. Another discussion of diquarks as building blocks for hadrons is in \cite{Friedmann:2014qpa}.}

\section{Basic theoretical model} \label{sec:basic}
As explained in the introduction our theoretical model is not the holographic stringy model, but a model in flat space-time that incorporates two ingredients motivated by the holographic picture. Namely, they are massive particles at the ends of the strings and a baryonic vertex that connects $N_c=3$ strings.

		\subsection{Classical rotating string with massive endpoints}
		We describe the string with massive endpoints (in flat space-time) by adding to the Nambu-Goto action,
		\be S_{NG} = -T\int\!\!{d\tau d\sigma \sqrt{-h}} \ee
		\[ h_{\alpha\beta} \equiv \eta_{\mu\nu} \del_\alpha X^\mu \del_\beta X^\nu \]
		a boundary term - the action of a massive chargeless point particle
		\be S_{pp} = -m\int\!\! d\tau \sqrt{-\dot{X}^2} \ee
		\[ \dot{X}^\mu \equiv \del_\tau X^\mu \]
		at both ends. There can be different masses at the ends, but here we assume, for simplicity's sake, that they are equal. We also define \(\sigma = \pm l\) to be the boundaries, with \(l\) an arbitrary constant with dimensions of length.

		The variation of the action gives the bulk equations of motion
		\be \del_\alpha\left(\sqrt{-h}h^{\alpha\beta}\del_\beta X^\mu\right) = 0 \label{eq:bulk} \ee
		and at the two boundaries the condition
		\be T\sqrt{-h}\del^\sigma X^\mu \pm m\del_\tau\left(\frac{\dot{X}^\mu}{\sqrt{-\dot{X}^2}}\right) = 0 \label{eq:boundary} \ee
		
		It can be shown that the rotating configuration
		\be X^0 = \tau, X^1 = R(\sigma)\cos(\omega\tau), X^2 = R(\sigma)\sin(\omega\tau) \ee
		solves the bulk equations \eqref{eq:bulk} for any choice of \(R(\sigma)\). We will use the simplest choice, \(R(\sigma) = \sigma\), from here on.\footnote{Another common choice is $X^0=\tau, x^1=\sin(\sigma) \cos(\omega\tau), X^2= \sin(\sigma)\sin(\omega\tau)$.} Eq. \eqref{eq:boundary} reduces then to the condition that at the boundary,
		\be \frac{T}{\gamma} = \gamma m \omega^2 l \label{eq:boundaryRot}\ee
		with \(\gamma^{-1} \equiv \sqrt{1-\omega^2 l^2}\).\footnote{Notice that in addition to the usual term $\gamma m$ for the mass, the tension that balances the ``centrifugal force" is $\frac{T}{\gamma}$.}

		We then derive the Noether charges associated with the Poincar\'e invariance of the action, which include contributions both from the string and from the point particles at the boundaries. Calculating them for the rotating solution, we arrive at the expressions for the energy and angular momentum associated with this configuration:
	\be	E = -p_0 = 2\gamma m + T \int_{-l}^l\!\frac{d\sigma}{\sqrt{1-\omega^2\sigma^2}} \ee
	\be J = J^{12} = 2\gamma m \omega l^2 + T \omega \int_{-l}^l\! \frac{\sigma^2 d\sigma}{\sqrt{1-\omega^2\sigma^2}} \ee

		Solving the integrals, and defining \(q \equiv \omega l\) - physically, the endpoint velocity - we write the expressions in the form
		\be E = \frac{2m}{\sqrt{1-q^2}} + 2Tl\frac{\arcsin(q)}{q} \ee
		\be J = 2m l \frac{q}{\sqrt{1-q^2}} + Tl^2\left(\frac{\arcsin(q)-q\sqrt{1-q^2}}{q^2}\right) \ee
		The terms proportional to \(m\) are the contributions from the endpoint masses and the term proportional to \(T\) is the string's contribution. These expressions are supplemented by condition \eqref{eq:boundaryRot}, which we rewrite as
		\be Tl = \frac{mq^2}{1-q^2} \label{eq:Tm}\ee
		This last equation can be used to eliminate one of the parameters \(l, m , T,\) and \(q\) from \(J\) and \(E\). Eliminating the string length from the equations we arrive at the final form
		\be E = 2m \left(\frac{q\arcsin(q)+\sqrt{1-q^2}}{1-q^2}\right) \label{eq:massiveE} \ee
		\be J = \frac{m^2}{T}\frac{q^2}{(1-q^2)^2}\left(\arcsin(q)+q\sqrt{1-q^2}\right) \label{eq:massiveJ} \ee

		These two equations are what define the Regge trajectories of the string with massive endpoints. They determine the functional dependence of \(J\) on \(E\), where they are related through the parameter \(0 \leq q < 1\) (\(q = 1\) when \(m = 0\)). Since the expressions are hard to make sense of in their current form, we turn to two opposing limits - the low mass and the high mass approximations.

			In the low mass limit where the endpoints move at a speed close to the speed of light, so \(q \rightarrow 1\), we have an expansion in \((m/E)\):
		\be J = \alp E^2\left(1-\frac{8\sqrt{\pi}}{3}\left(\frac{m}{E}\right)^{3/2} + \frac{2\sqrt{\pi^3}}{5}\left(\frac{m}{E}\right)^{5/2} + \cdots\right) \label{eq:lowMass}\ee
		from which we can easily see that the linear Regge behavior is restored in the limit \(m\rightarrow 0\), and that the first correction is proportional to \(\sqrt{E}\). The Regge slope \(\alp\) is related to the string tension by \(\alp = (2\pi T)^{-1}\).

		The high mass limit, \(q \rightarrow 0\), holds when \((E-2m)/2m \ll 1\). Then the expansion is
		\be J = \frac{4\pi}{3\sqrt{3}}\alp m^{1/2} (E-2m)^{3/2} + \frac{7\pi}{54\sqrt{3}} \alp m^{-1/2} (E-2m)^{5/2}
			+ \cdots \label{eq:highMass} \ee
		
	\subsubsection{Generalizations: Different masses, Y-shape model, and central mass} \label{sec:Generalizations}
		The generalization of the symmetric model to the case where there are two different masses is simple. The angular momentum and energy were calculated by summing two equal contributions - from two halves of the string (\(\sigma\in(-l,0)\) and \(\sigma\in(0,l)\)) and two identical point particles. To generalize this we simply replace the factor of two with a sum over two different, but similar, contributions.
		
		The two masses would still rotate with the same angular velocity, \(\omega\), and the string tension remains the same for both string segments. The difference between the two contributions stems from different endpoint velocities, \(q_i\), and the different radii of rotation, related to the velocities by \(\omega l_i = q_i\).
		
		The energy and angular momentum in this case are, then,
		\be E = \sum_{i=1,2}m_i\left(\frac{q_i\arcsin(q_i)+\sqrt{1-q_i^2}}{1-q_i^2}\right) \label{eq:massGenE} \ee
		\be J = \sum_{i=1,2}\pi\alp m_i^2\frac{q_i^2}{(1-q_i^2)^2}\left(\arcsin(q_i)+q_i\sqrt{1-q_i^2}\right) \label{eq:massGenJ} \ee
		The velocities \(q_1\) and \(q_2\) can be related using the boundary condition \eqref{eq:boundaryRot}, from which we have
		\be \frac{T}{\omega} = m_1\frac{q_1}{1-q_1^2} = m_2\frac{q_2}{1-q_2^2} \label{eq:boundaryGen}\ee
		With \(q_1\) and \(q_2\) thus related, the massive Regge trajectory is obtained from the parametric curve
			\be E = E\left(q_1,q_2(q_1)\right) \qquad J = J\left(q_1,q_2(q_1)\right) \ee
		where \(0 \leq q_1 < 1\)
		
		The equations of motion for the three segments of the Y-shaped string are unchanged from the simple straight string. The only difference is an added boundary condition at the point where the three strings connect. While this added condition is important when analyzing the stability of the model, as done e.g. in \cite{'tHooft:2004he}, it holds trivially in the unperturbed rotating solution. Therefore the only adjustment we need make to eqs. \eqref{eq:massGenE} and \eqref{eq:massGenJ} to get to the trajectories of the Y-shape model is in the summation index \(i\) - we now sum over three contributions from three string segments and three endpoint masses, so \(i = 1,2,3\).
		
		In the case where we have three identical end point masses we have
		\be E = 3m\left(\frac{q\arcsin(q)+\sqrt{1-q^2}}{1-q^2}\right) \label{eq:massThreeE} \ee
		\be J = \frac{3}{2}\frac{m^2}{T}\frac{q^2}{(1-q^2)^2}\left(\arcsin(q)+q\sqrt{1-q^2}\right) \label{eq:massThreeJ} \ee
		This model is completely equivalent (in terms of the Regge trajectory) with a single string model with the same total mass at the endpoints and with an effective higher string tension. That is to say, a Y-shaped string with three identical masses \(m_Y\) and the string tension \(T_Y\) has the same exact Regge behavior as a string with two masses \(m_1 = m_2 = 3m_Y/2\) and the tension \(T\), provided \(T_Y = \frac{3}{2}T\). In particular, in the massless case the Y-shape model yields the result
		\be J = \frac{2}{3}\alp E^2 \label{eq:alpY} \ee
		
		Another generalization would be to add a central mass to the string around which the string rotates. This mass would be stationary so its only contribution would be a constant shift in the energy, which amounts to the modification \(E \rightarrow E - m_{bv}\) in previous equations.\footnote{The name \(m_{bv}\) stems from a possible identification of such a central mass with the presence of a baryonic vertex encountered in the holographic models.}
	
\section{Fitting models} \label{sec:models}
	\subsection{Rotating string model}
		We define the \emph{linear} fit by
			\be J + n = \alp E^2 + a \label{eq:linear} \ee
		where the fitting parameters are the \emph{slope} \alp and the \emph{intercept} \(a\).
				
		For the \emph{massive} fit, we use the general expressions for the mass and angular momentum of the rotating string, eqs. \eqref{eq:massGenE} and \eqref{eq:massGenJ}, for the case of two different masses, and we add to them, by hand, an intercept and an extrapolated \(n\) dependence, assuming the same replacement of \(J \rightarrow J + n - a\).
		\be E = \sum_{i = 1,2}m_i\left(\frac{q_i\arcsin(q_i)+\sqrt{1-q_i^2}}{1-q_i^2}\right) \label{eq:massFitE} \ee
	\be J + n = a + \sum_{i=1,2}\pi\alp m_i^2\frac{q_i^2}{(1-q_i^2)^2}\left(\arcsin(q_i)+q_i\sqrt{1-q_i^2}\right) \label{eq:massFitJ} \ee
		With the relation between \(q_1\) and \(q_2\) as in eq. \eqref{eq:boundaryGen}:
		\be \frac{T}{\omega} = m_1\frac{q_1}{1-q_1^2} = m_2\frac{q_2}{1-q_2^2} \label{eq:boundaryTwo}\ee
		
		With the two additions of \(n\) and \(a\), the two equations reduce to that of the linear fit in \eqref{eq:linear} in the limit where both masses are zero.
		
		Now the fitting parameters are \(a\) and \(\alp\) as before, as well as the the two endpoint masses \(m_1\) and \(m_2\).
		
		\subsubsection{Rotating Y-shape with central mass}
		We also examine fits using the Y-shape model described in section \ref{sec:Generalizations}, as well as the model where the string rotates about a stationary mass, \(m_{bv}\).
		
		The Y-shape fit uses the same expressions as the massive fit with the summation changed from \(i=1,2\) to \(i = 1,2,3\). The central mass \(m_{bv}\) may be inserted to the fits by the replacement \(E \rightarrow E - m_{bv}\) in eq. \eqref{eq:massFitE}, the same way we have inserted the intercept \(a\) to \(J\). Explicitly, the expressions then are
		\be E = m_{bv} + \sum_{i = 1}^3m_i\left(\frac{q_i\arcsin(q_i)+\sqrt{1-q_i^2}}{1-q_i^2}\right) \label{eq:massFitYE} \ee
	\be J + n = a + \sum_{i=1}^3\pi\alp m_i^2\frac{q_i^2}{(1-q_i^2)^2}\left(\arcsin(q_i)+q_i\sqrt{1-q_i^2}\right) \label{eq:massFitYJ} \ee
		The relation between the three velocities \(q_i\) is again through
		\be \frac{T}{\omega} = m_i\frac{q_i}{1-q_i^2} \ee
		for \(i = 1,2,3\). In the limit where all endpoint masses are zero, these equations reduce to
		\be J + n - a = \frac{2}{3}\alp(E-m_{bv})^2 \ee
		
\subsection{Fitting procedure}
			
		We measure the quality of a fit by the dimensionless quantity \(\chi^2\), which we define, as we did in \cite{Sonnenschein:2014jwa}, by
				\be \chi^2 = \frac{1}{N-1}\sum_i\left(\frac{M_i^2-E_i^2}{M_i^2}\right)^2 \label{eq:chi_def} \ee
			\(M_i\) and \(E_i\) are, respectively, the measured and calculated value of the mass of the \(i\)-th particle, and \(N\) the number of points in the trajectory.
			
\section{Fit results} \label{sec:results}
	This section offers a discussion of the fit results for the baryons. We start by briefly discussing the results using the Y-shaped string model for the baryon, and the effects of including a massive baryonic vertex at the center of mass, via the replacement \(M \rightarrow M-m_{bv}\). The results prove to be against these options, so the rest of the section discusses the results when using the quark-diquark model for the baryons. This means we describe the baryons simply as a single string with two masses at its endpoints.
	
We separate the results into three sections, one for the light quark baryons, the next for strange baryons, and the third for charmed baryons. In the light baryon section we also examine the radial trajectories of the \(N\) and \(\Delta\) baryons. The rest of the sections have trajectories only in the angular momentum plane \((J,M^2)\).
	
	The detailed individual trajectory fits and the specification of all the states used are found in appendix \ref{app:individualb}. The experimental data used in this paper is taken from the Particle Data Group's Review of Particle Physics\cite{PDG:2012}.

	\subsection{Y-shape and central mass}
	As explained in section \ref{sec:Generalizations}, the Y-shaped string model is equivalent in terms of the Regge trajectories to a single string model with a higher effective string tension. In the picture we have before adding endpoint masses, we may assume (as a phenomenological model) linear trajectories for both the meson and baryon trajectories,
	\be J = \alp M^2 + a \ee
	with different slopes, \(\alp\!_b\) and \(\alp\!_m\) for the baryons and mesons, respectively. Now, the assumption that the baryons are Y-shaped strings while the mesons are straight single strings, and that there is a single universal string tension, would lead us to expect the relation
	\be \alp\!_b = \frac{2}{3}\alp\!_m \ee
	between the baryon and meson Regge slopes. When we fit the data we see no such relation. What we see in our results is that in fact, the meson and baryon slopes are very similar - \(\alp\!_b \approx \alp_m\) - supporting the same single string model for the baryons that was used for the mesons. This also excludes the triangle-shaped closed string baryon, which we have not analyzed in detail but predicts an effective slope \(\alp\!_b\) of between \(\frac{3}{8}\alp\!_m\) and \(\frac{1}{2}\alp\!_m\), depending on the type of solution \cite{Sharov:1998hi}.
	
Our addition of endpoint masses does not change this picture, as we would still need to see a similar relation between the baryon and meson slopes, with the baryon slope being consistently lower. The slopes obtained from the baryon trajectory fits, which we present in the next sections, still remain too high for the Y-shape model of the baryon to be consistent with experimental data. In many cases the baryon slope is actually higher than the meson slope.
	
As for the baryonic vertex mass, the assumption that there is a central mass that contributes to the total mass of a state but not to the angular momentum was also found to be unsupported by the data, neither in the Y-shape model nor in the straight string model.\footnote{The two-string model discussed in section \ref{sec:stability} with one quark at the center of mass has the same slope as the quark-diquark model, but we exclude it here (after excluding it on grounds of its instability in section \ref{sec:stability}) because we see no evidence, in any trajectory, of a massive central point.} We plot one of these fits in a following section (bottom-right plot in figure (\ref{fig:chi_b_light})). This does not rule out the presence of a mass due to a holographic baryonic vertex, but means it is most likely located at the string endpoints and not at its center.
	
	With these results in mind, we continue to present our fits using only the single string model with two masses at its endpoints.
	
	\subsection{Symmetric vs. imbalanced string} \label{sec:barsym}
\begin{figure}[t!] \centering
						\includegraphics[natwidth=1200bp, natheight=900bp, width=.44\textwidth]{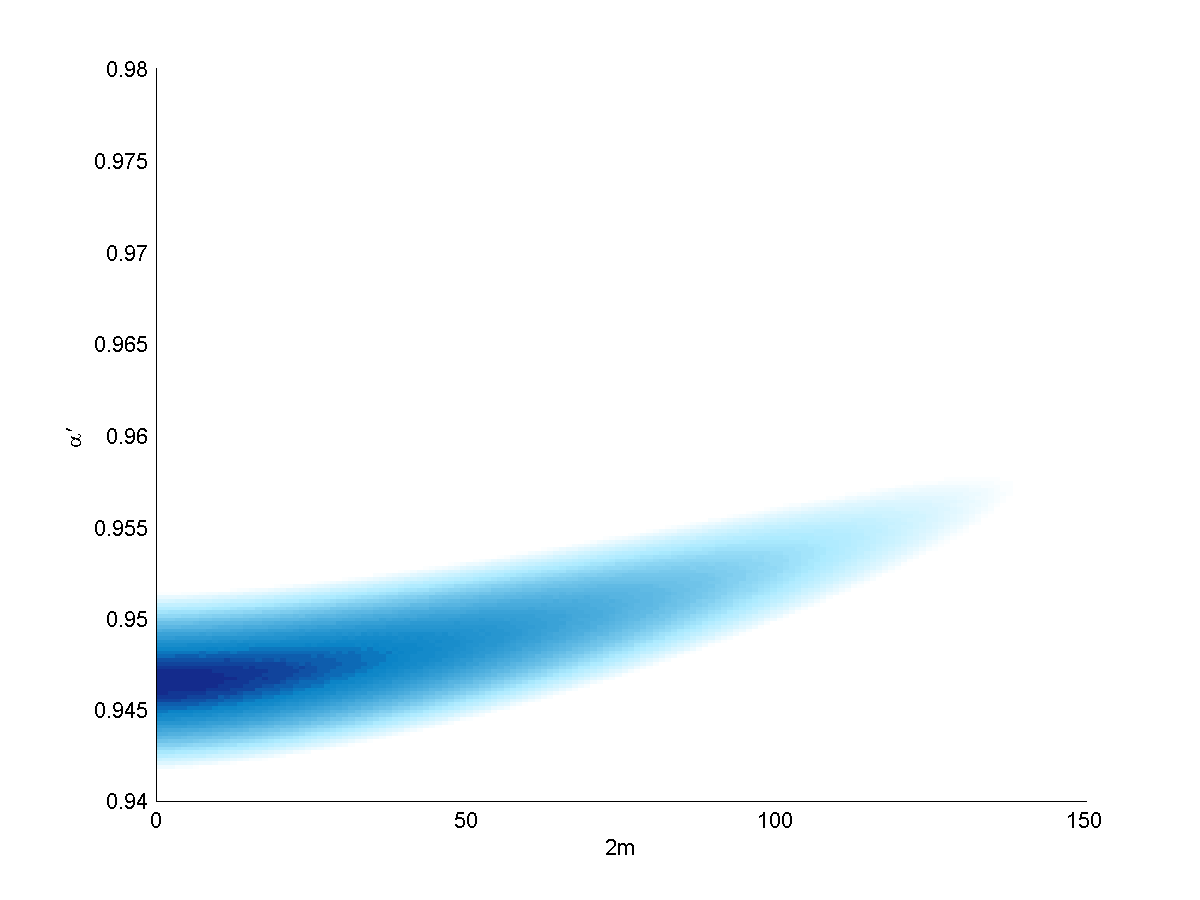}	 \hfill
						\includegraphics[natwidth=1200bp, natheight=900bp, width=.44\textwidth]{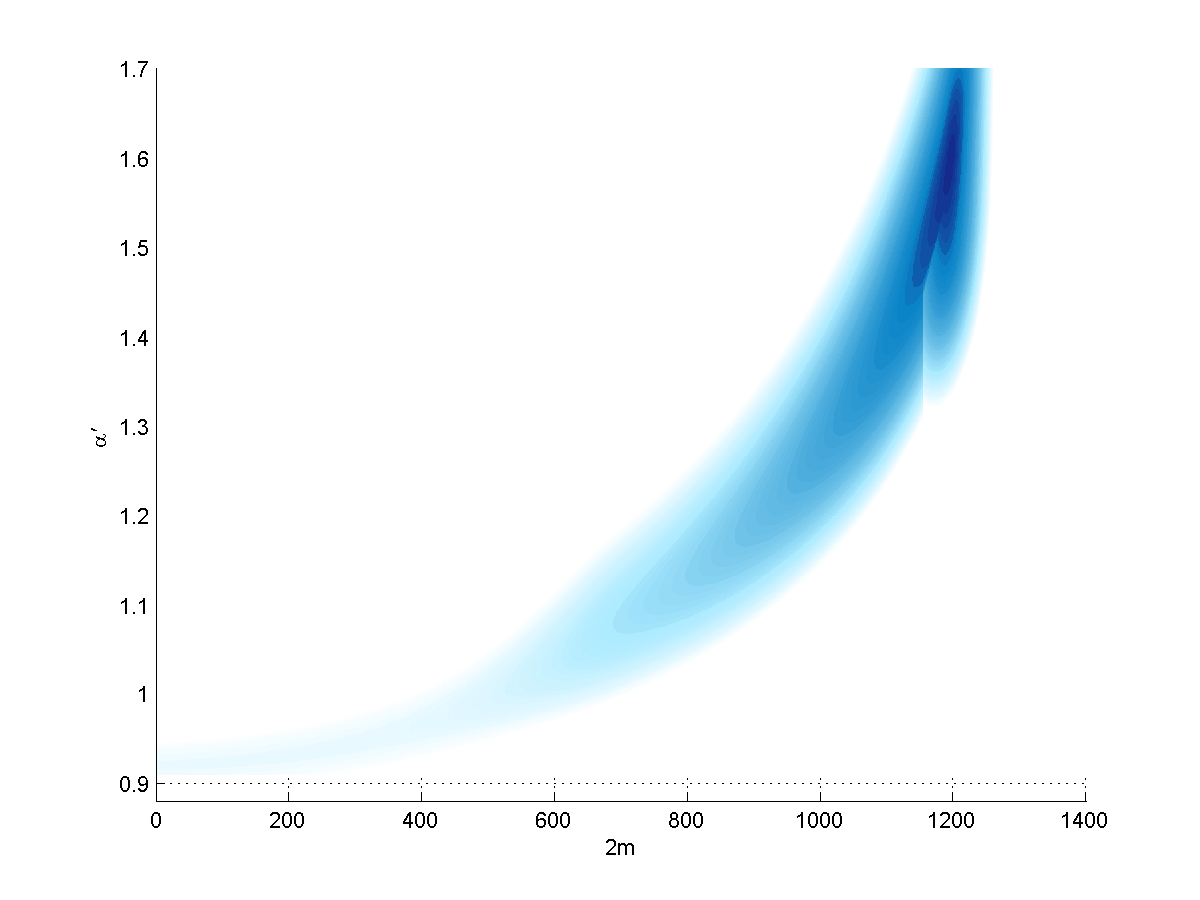} \\
						\includegraphics[natwidth=1200bp, natheight=900bp, width=.44\textwidth]{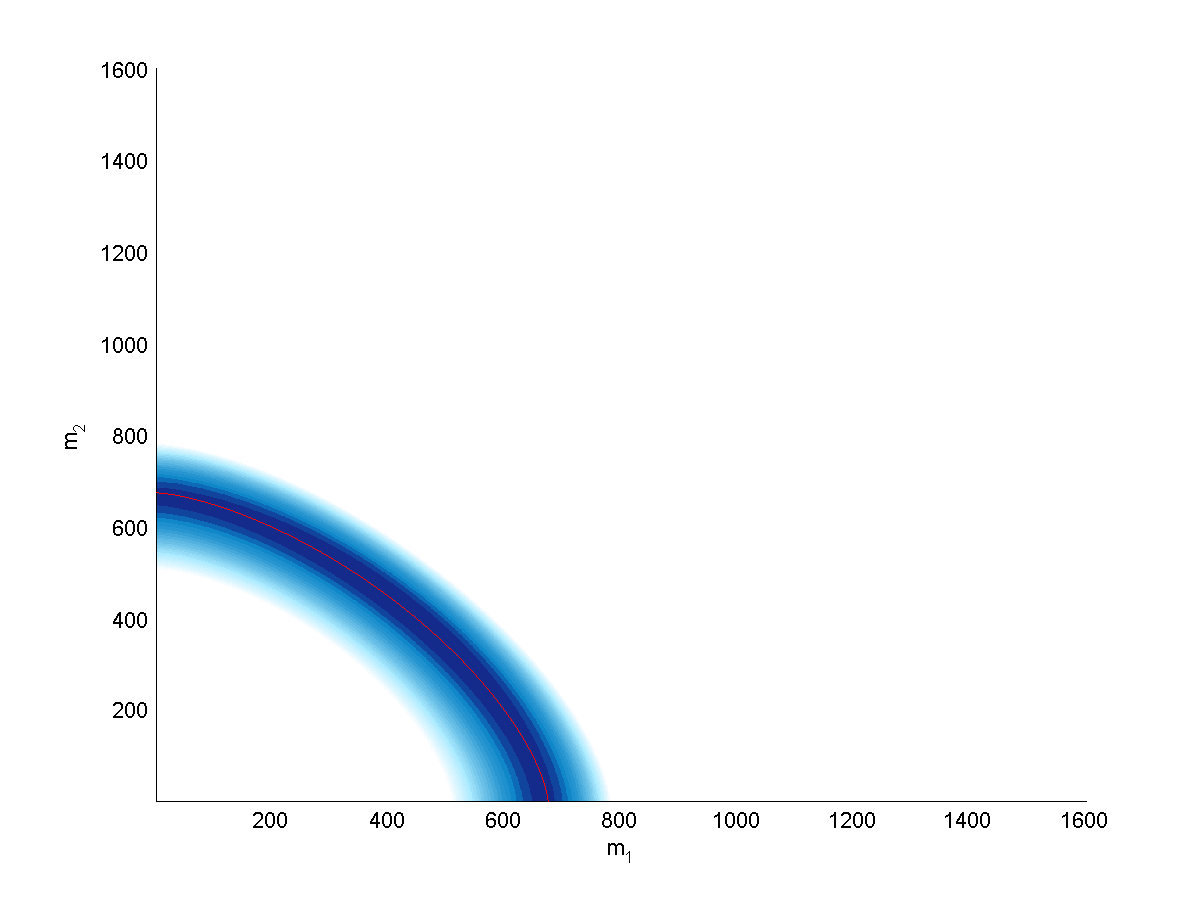}	 \hfill
						\includegraphics[natwidth=1200bp, natheight=900bp, width=.44\textwidth]{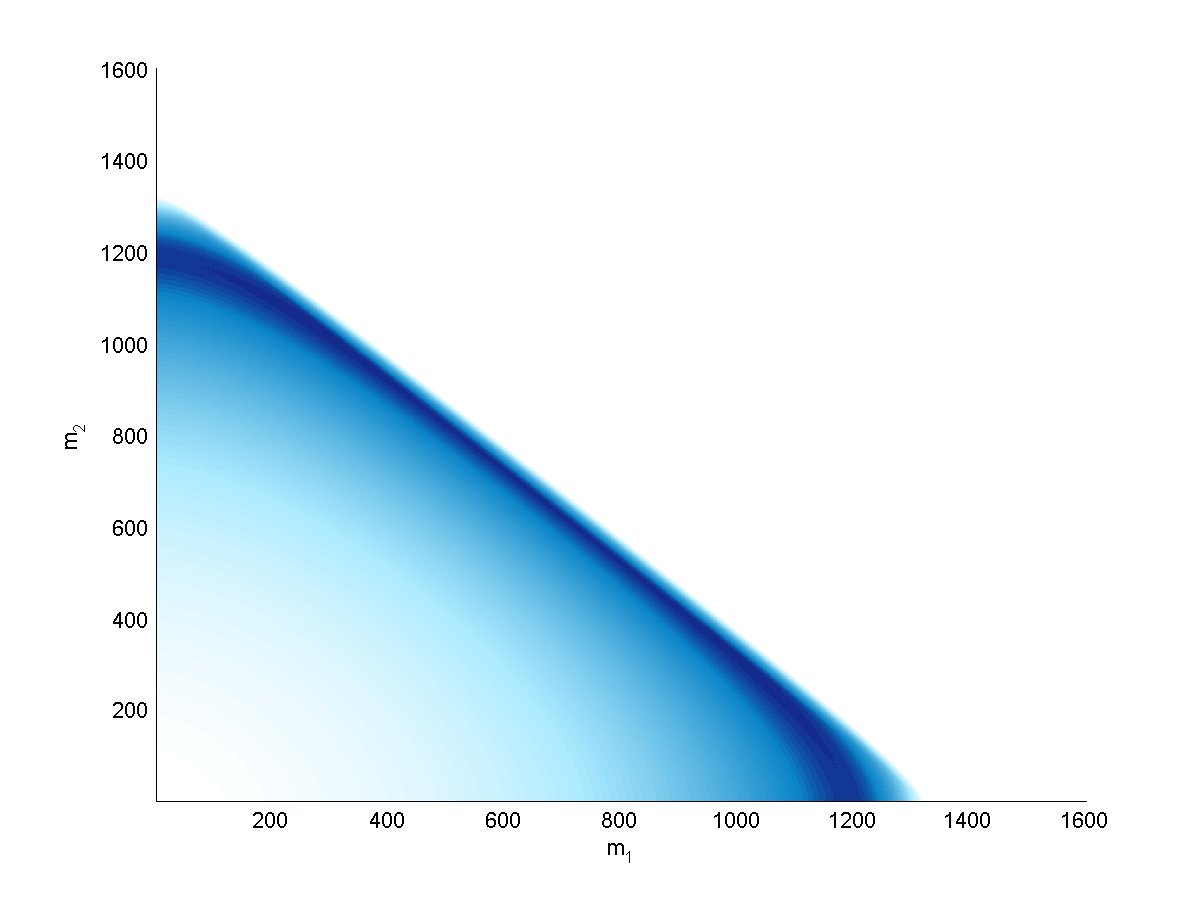} \\
						\caption{\label{fig:chi_b_s} The strange baryon trajectory fits. Top-left: \(\chi^2\) vs. \((\alp,2m)\) for the \(\Lambda\) trajectory. Top-right: same for the \(\Sigma\). Bottom: \(\chi^2\) vs. \((m_1,m_2\)) for the \(\Xi\) trajectory, for \(\alp = 0.950\) (left) and adjustable \(\alp\) (right). The red curve in the bottom-left plot is \(m_1^{3/2}+m_2^{3/2} = 2\times(425)^{3/2}\).}
	\end{figure}

	Now we turn to the different mass configurations in the single string model. As mentioned in the last subsection, there is no evidence to support a mass located at the center of the string. To understand the structure of the baryon we would like to be able to tell how the mass is distributed between the two endpoints, but this is information we cannot gather from the Regge trajectory fits.
	In the low mass approximation for the single string, the leading order correction is proportional to \(\alp(m_1^{3/2}+m_2^{3/2})\sqrt{E}\). This is the generalization of the low mass expansion of eq. \eqref{eq:lowMass}. Therefore, for small masses - \((m_1+m_2)/E \ll 1\) - we cannot distinguish from the Regge trajectory fits alone between different configurations with equal \(m_1^{3/2}+m_2^{3/2}\). There are of course higher order corrections, but our fits are not sensitive enough to them, and in practice, we see that fits with \(m_1^{3/2}+m_2^{3/2} = Const.\) are nearly equivalent even for masses of a few hundred MeV (when using a typical value of the slope, in the neighborhood of \(0.9\) GeV\(^{-2}\)). An example of this is in the bottom-left plot of figure (\ref{fig:chi_b_s}).
	
	When expanding \(J\) in \(E\) for two heavy masses, the combination in the leading term would be \(m_1 + m_2\). In the mid range that cannot be described accurately by either expansion there is a transition between the two different type of curves. This can be best seen in figure (\ref{fig:chi_b_c}), which shows the fits of the charmed \(\Lambda_c\) baryon. For higher masses, the variation of the intercept along these curves is big enough to be easily measured, but as the intercept takes reasonable values for both symmetric and asymmetric massive fits we cannot use this measurement to decide between the two configurations. We do not have accurate predictions for the intercept, and neither will we find emerging in our fit results remarkably consistent results for it. To give some numbers, the symmetric fit of the \(\Lambda_c\) with \(2m = 2010\) has \(a = 0.09\), while the asymmetric configuration which gives a comparable fit has \(m_1 = 1720\), \(m_2 = 90\) and \(a = -0.13\). The slope is roughly equal between the two fits, \(1.13\) GeV\(^{-2}\) for the former and \(1.22\) GeV\(^{-2}\) for the latter. Since we have no reason to prefer \(a = 0.09\) over \(a = -0.13\) we cannot use this information.
	
	In both cases the symmetric fit where \(m_1 = m_2 = m\) gives an indication of the total mass we can add to the endpoints for a good fit of a given trajectory. The best fitting masses are on a curve in the \((m_1,m_2)\) plane. The choice \(m_1 = m_2\) maximizes the total mass \(m_1 + m_2\), whether the masses are light or heavy.
	
	It should also be noted that for the trajectories we analyze we either have fits with low masses, where the \(m^{3/2}\) approximation is valid, or trajectories with only 3 data points where we would by default expect the optimum to be located on a curve in the \((m_1,m_2)\) plane, seeing how there are four fitting parameters in total.
	
	While the presentation in the following sections of the results is for the symmetric fit, this does not mean that we have found it is actually preferred by the data. The symmetric fit tells us whether there is a preference for non-zero endpoint masses or not, and allows us to obtain the values of the slope for a given total endpoints mass.
	
	If we assume \(\chi^2\) is constant along curves of the form \(m_1^{3/2}+m_2^{3/2} = Const.\), then there is not much difference between the symmetric case and the case \(m_1 = 2m_2\) when considering the total sum \(m_1 + m_2\). So for a given symmetric fit with \(2m\), the configuration with \((m_1,m_2) = (\frac{2}{3}m,\frac{4}{3}m)\) will be almost equivalent to the one with \(m_1 = m_2 = m\) (and this is true at both high and low masses). If one of the masses is zero, say \(m_2 = 0\), then as a rule of thumb one can take \(m_1 = 2^{2/3}m \approx 0.8\times2m\) to move from the symmetric fit result to the totally imbalanced mass configuration.
	
	\subsection{Light quark baryons}
	
	\begin{figure}[t!] \centering
						\includegraphics[natwidth=1200bp, natheight=900bp, width=.44\textwidth]{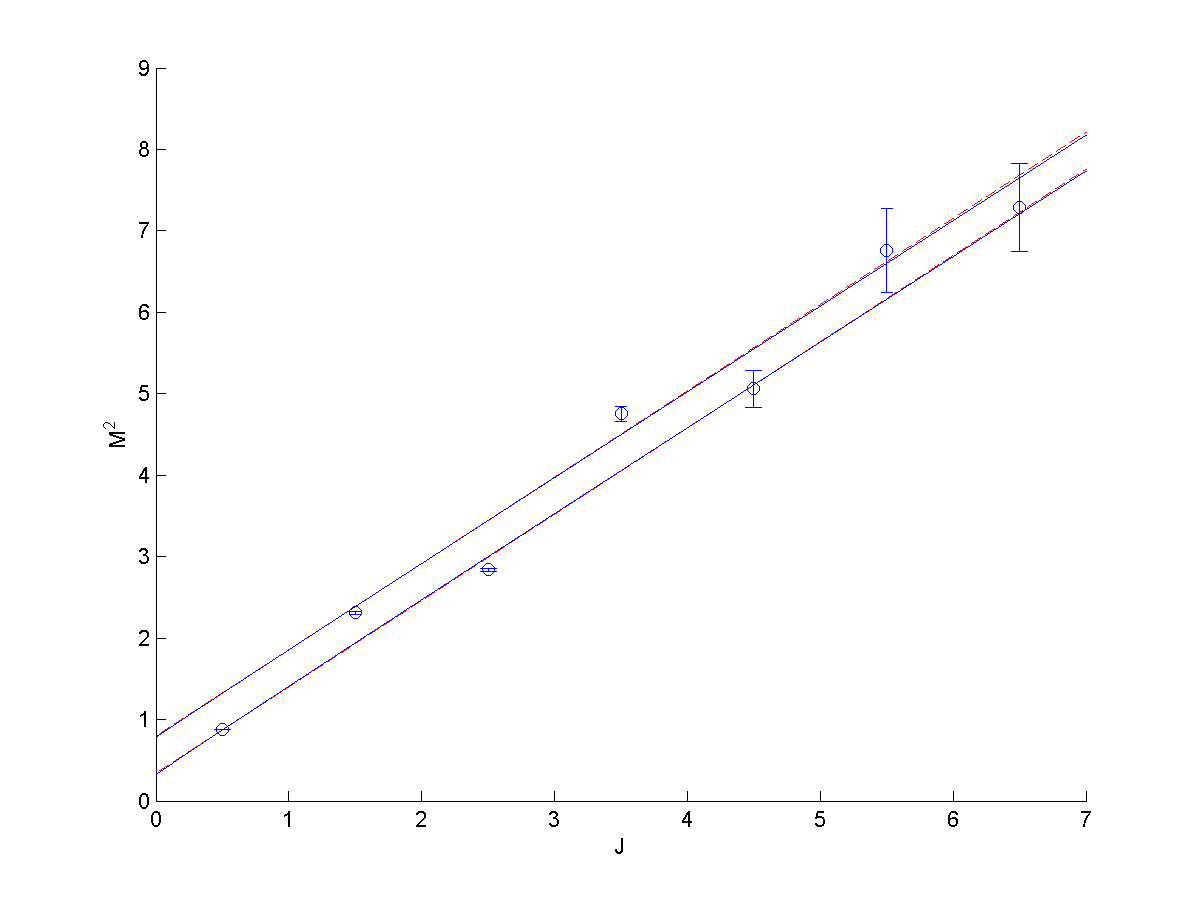}	 \hfill
						\includegraphics[natwidth=1200bp, natheight=900bp, width=.44\textwidth]{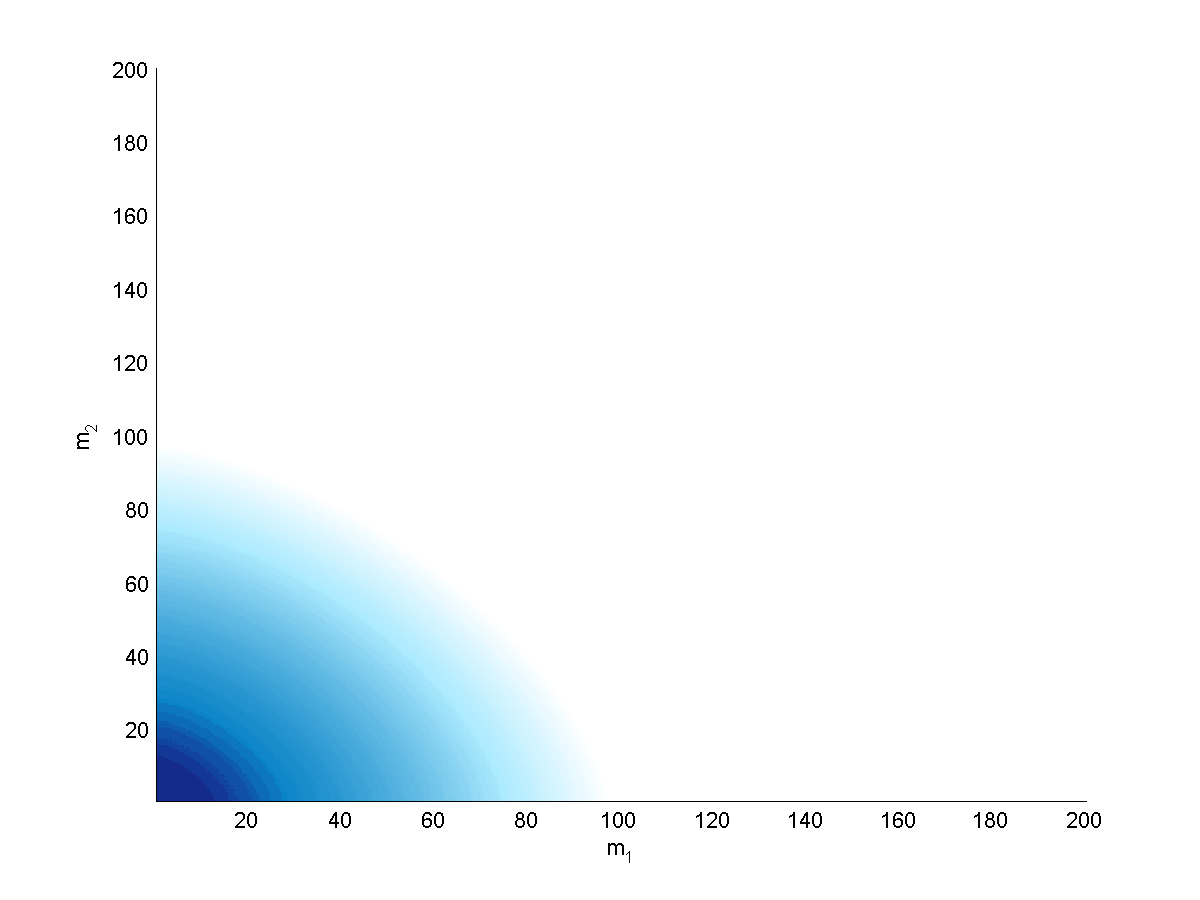} \\
						\includegraphics[natwidth=1200bp, natheight=900bp, width=.44\textwidth]{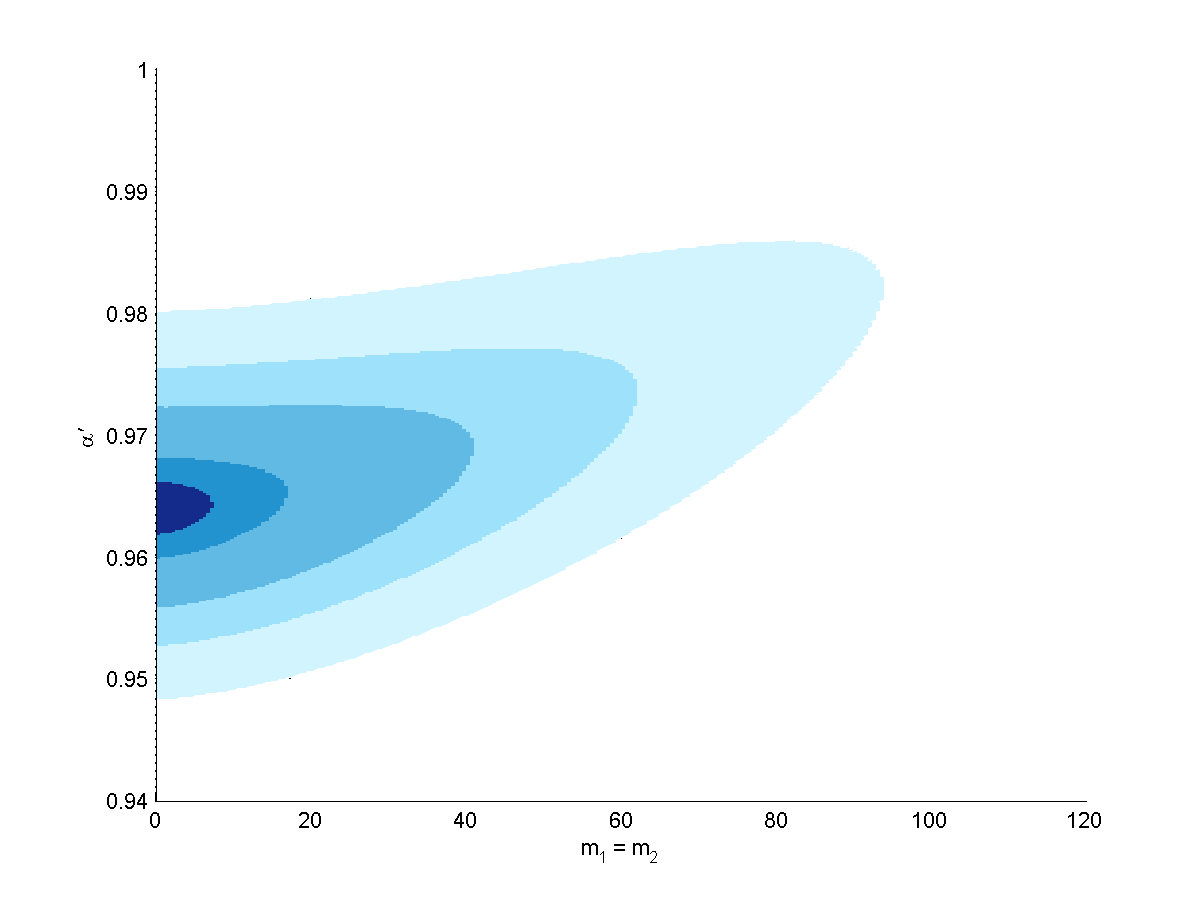}	 \hfill
						\includegraphics[natwidth=1200bp, natheight=900bp, width=.44\textwidth]{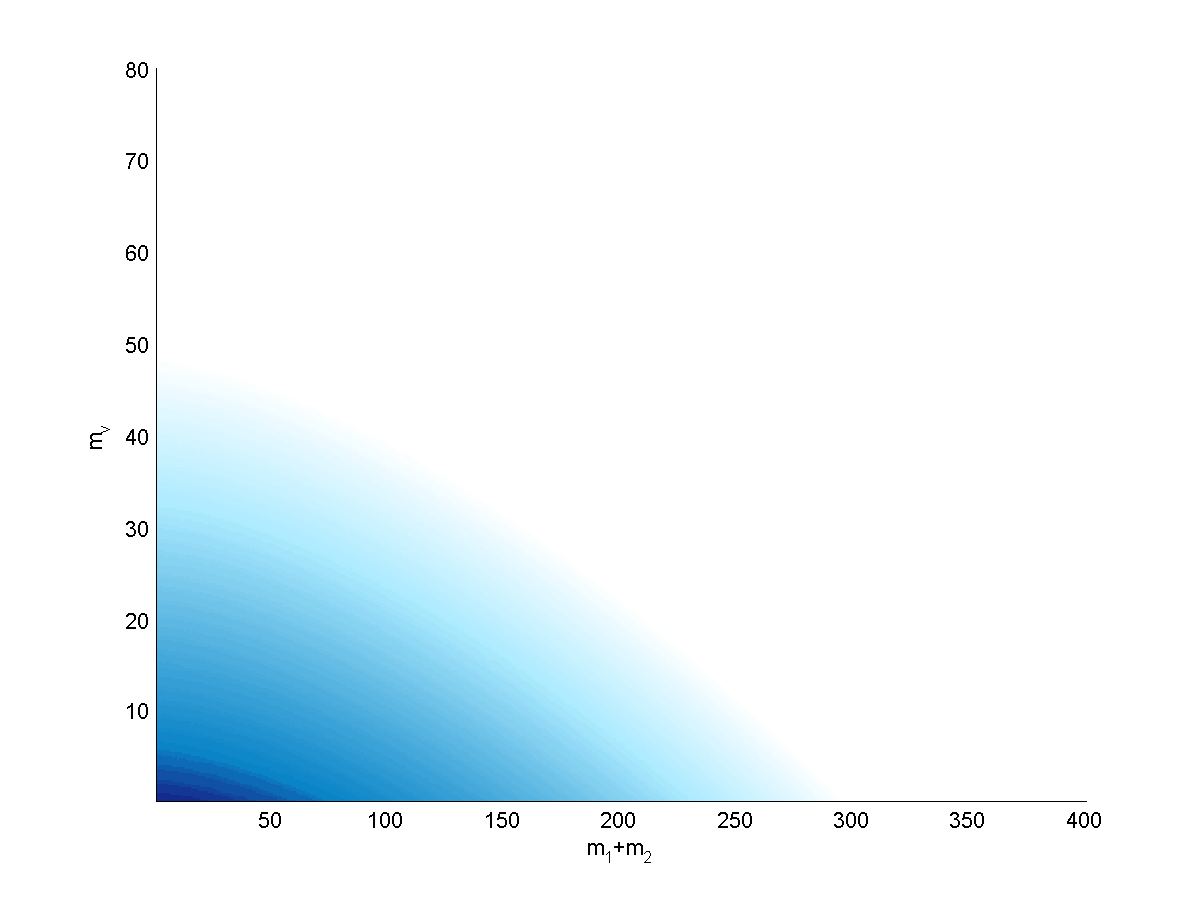} \\
						\caption{\label{fig:chi_b_light} The light baryon trajectory fits. Top-left: The \(N\) trajectory and fits, showing the even/odd effect. Top-right: \(\chi^2\) vs. the two masses for the \(N\) trajectory. Bottom-left: \(\chi^2\) vs. \((\alp,2m\)) for the \(N\). Bottom-right: \(\chi^2\) vs. \(2m\) and a central vertex mass \(m_{bv}\) for the \(\Delta\) trajectory.  The \(\chi^2\) plots are for fits to the \((J,M^2)\) trajectories, and use only the even \(L\) states. The darkest areas in the \(\chi^2\) plots have \(\chi_m^2/\chi^2_l < 1\), lightest areas are \(\chi_m^2/\chi_l^2 < 1.1\).}
	\end{figure}
	
		In the light baryon sector, we look at the \(N\) and \(\Delta\) resonances.
		\subsubsection{Trajectories in the \texorpdfstring{$(J,M^2)$}{(J,M2)} plane}
		One of the most interesting features of the baryon Regge trajectories is the splitting of the trajectories of even and odd \(L\) states. The states with even and odd orbital angular momentum do not lie on one single trajectory, but on two parallel linear trajectories, the odd \(L\) states being higher in mass and lying above the trajectory formed by the even states. The plot in figure (\ref{fig:chi_b_light}) shows this effect for the \((J,M^2)\) trajectory of the \(N\).
		
		In our analysis, we fit the even and odd trajectories together, with the same endpoint masses and slope, and allow the intercept to carry the difference between the even and odd states.
		
		In this way, we get that the \(N\) trajectory is best fitted with a slope of around \(0.95\) GeV\(^{-2}\), and that the linear fit is optimal. Only small masses, up to a total mass of \(2m = 170\) MeV, are allowed.\footnote{Masses in what we call the ``allowed'' range give a value of \(\chi^2\) that is within 10\(\%\) of its optimal value for that specific trajectory.} Trying a fit using only the four highest \(J\) states (two even and two odd), we achieve a weaker \(\chi^2\) dependence on the mass, and we can add a total mass of up to \(640\) MeV, with the slope being near \(1\) GeV\(^{-2}\) for the highest masses.
		
		The \(\Delta\) is also best fitted by the linear, massless, trajectory, but it allows for higher masses. The maximum for it is \(450\) MeV. The slope, for a given mass, is lower than that of the \(N\). It is between 0.9 GeV\(^{-2}\) for the linear fit, and 0.97 GeV\(^{-2}\) for the maximal massive fits.
		
		As for the even-odd effect, we quantify it by looking at the difference between the intercept obtained for the even \(L\) trajectory and the one obtained for the odd \(L\) trajectory. The magnitude of the even-odd splitting is higher for the \(\Delta\) than it is for the \(N\) baryons. For the \(\Delta\), the difference in the intercept is of almost one unit - \(a_e - a_o \approx 1\), while for the \(N\) it is less than half that: \(a_e - a_o \approx 0.45\). The difference in \(M^2\) is obtained by dividing by \(\alp\), so it is \(0.5\) GeV\(^{-2}\) for the \(N\) and \(1.1\) GeV\(^2\) for the \(\Delta\).
		
		\subsubsection{Trajectories in the \texorpdfstring{$(n,M^2)$}{(n,M2)} plane}
		The radial trajectories we analyze are also best fitted with small, or even zero, endpoint masses.
		
		For the \(\Delta\) we have three states with \(J^P = \jph{3}{+}\).	The slope is between 0.92 and 0.94 GeV\(^{-2}\) and the maximal allowed total mass is less than 200 MeV.
		
		For the \(N\) we use a total of fifteen states: four with \(J^P = \jph{1}{+}\) (the neutron/proton and higher resonances), three with \jph{3}{-}, and four pairs with other \(J^P\) assignments. They are all fitted with the same slope and mass. The results show a lower slope here, from 0.82 GeV\(^{-2}\) for the linear fit to \(0.85\) GeV\(^{-2}\) for the highest mass fit, this time with \(2m = 425\) MeV.
		
\subsection{Strange baryons}
	In the strange section there are several trajectories we analyze.
	
	The first is that of the \(\Lambda\). There are five states in this trajectory, enough for us to see that the even-odd effect is not present - or too weak to be noticeable. The linear fit, with \(\alp = 0.95\) GeV\(^{-2}\), is the optimal fit, and only small masses of the order of 60 MeV are allowed at each endpoint. Even if one of the masses is zero, the mass at the other end could not exceed 100 MeV. This is a puzzling result because the results of the meson fits, which will be compared in detail to the baryon fits in a later section, point toward a mass of \(200-400\) MeV for the \(s\) quark.
	
	On the other hand, the other strange baryon trajectories we examine point to very high masses, with correspondingly high values of the slope.	These are the results of the two trajectories of the \(\Sigma\) baryon we examine. The first has three states with \(J^P = \) \jph{1}{+}, \jph{3}{-}, and \jph{5}{+}. The second has the parity reversed (for a given value of \(J\)): \(J^P\) = \jph{3}{+}, \jph{5}{-}, and \jph{7}{+}.
	
	Since there are only three states per trajectory we cannot determine from the data alone whether or not there is an even-odd splitting effect present here (and this is the case with all following trajectories). Assuming that there is no even-odd splitting, the best fits have \(2m \approx 1200\) MeV and a slope of about \(1.4-1.5\) GeV\(^{-2}\). Assuming splitting, we find in the case of the \(\Sigma\) that the linear fit connecting the two even states has \(\alp \approx 0.9\) GeV\(^{-2}\).
	
	A third option, is fixing the slope at a more reasonable low value - we chose \(\alp = 0.95\) GeV\(^{-2}\) - and redoing the massive fits (with the assumption that there is no splitting). The best fits then for the \(\Sigma\) are at around \(2m = 500\) MeV, with the mass being somewhat higher in the trajectory beginning with of the \jph{1}{+} state. This is certainly the choice that is most consistent with previous results, as we can distribute the total mass so there is a mass of \(m_s \approx 400\) MeV at one end and up to 100 MeV at the other. The cost in \(\chi^2\) is fairly high: for the first trajectory \(\chi^2\) is approximately \(10^{-4}\) for the higher slope and ten times larger for \(\alp = 0.95\), while for the second \(\chi^2\) is almost zero for the high slope fit\footnote{This is often the case with three point trajectories, where we may find a choice of the parameters for which the trajectory passes through all three data points. This makes the error in the measurement hard to quantify.} and about \(5\ten{-4}\) for the fixed slope fit. In any case, the fits with \(\alp = 0.95\) GeV\(^{-2}\) and with the added masses have a better \(\chi^2\) than the linear massless fits.
	
	There is one more possible trajectory we examine, of the doubly strange \(\Xi\) baryon. The best fit overall is again with \(\alp \approx 1.5\) GeV\(^{-2}\), and at a somewhat higher mass of \(2m = 1320\). Fixing the slope at \(0.95\) GeV\(^{-2}\) results in \(2m = 850\) being optimal. This is again the best choice in terms of consistency - the total mass is exactly in the range we would expect to see where there are two \(s\) quarks present. In \(\chi^2\), the fit with the high slope has \(\chi^2 \approx 10^{-4}\) while the latter has \(\chi^2 \approx 4\ten{-4}\).
	
	\subsection{Charmed baryons}

	\begin{figure}[t!] \centering
						\includegraphics[natwidth=1200bp, natheight=900bp, width=.48\textwidth]{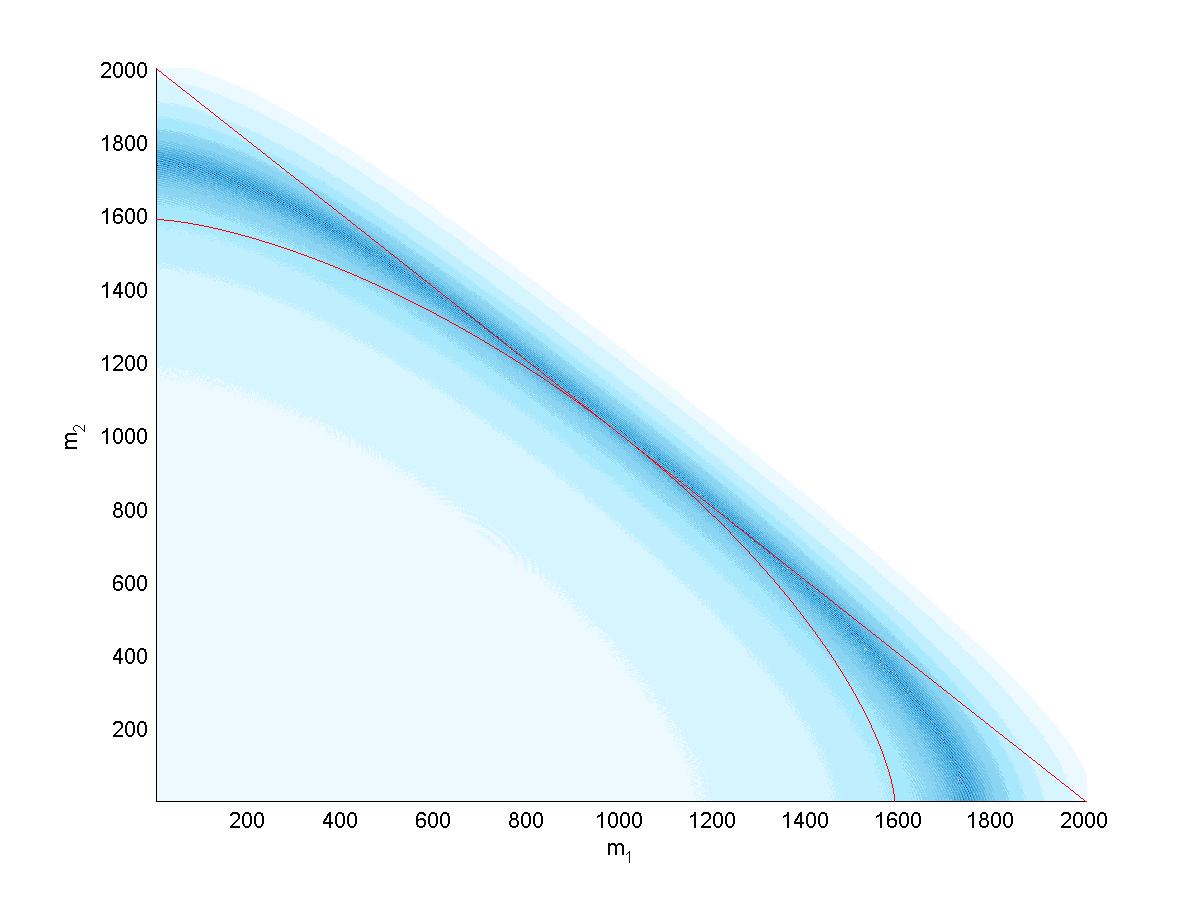}	 \hfill
						\includegraphics[natwidth=1200bp, natheight=900bp, width=.48\textwidth]{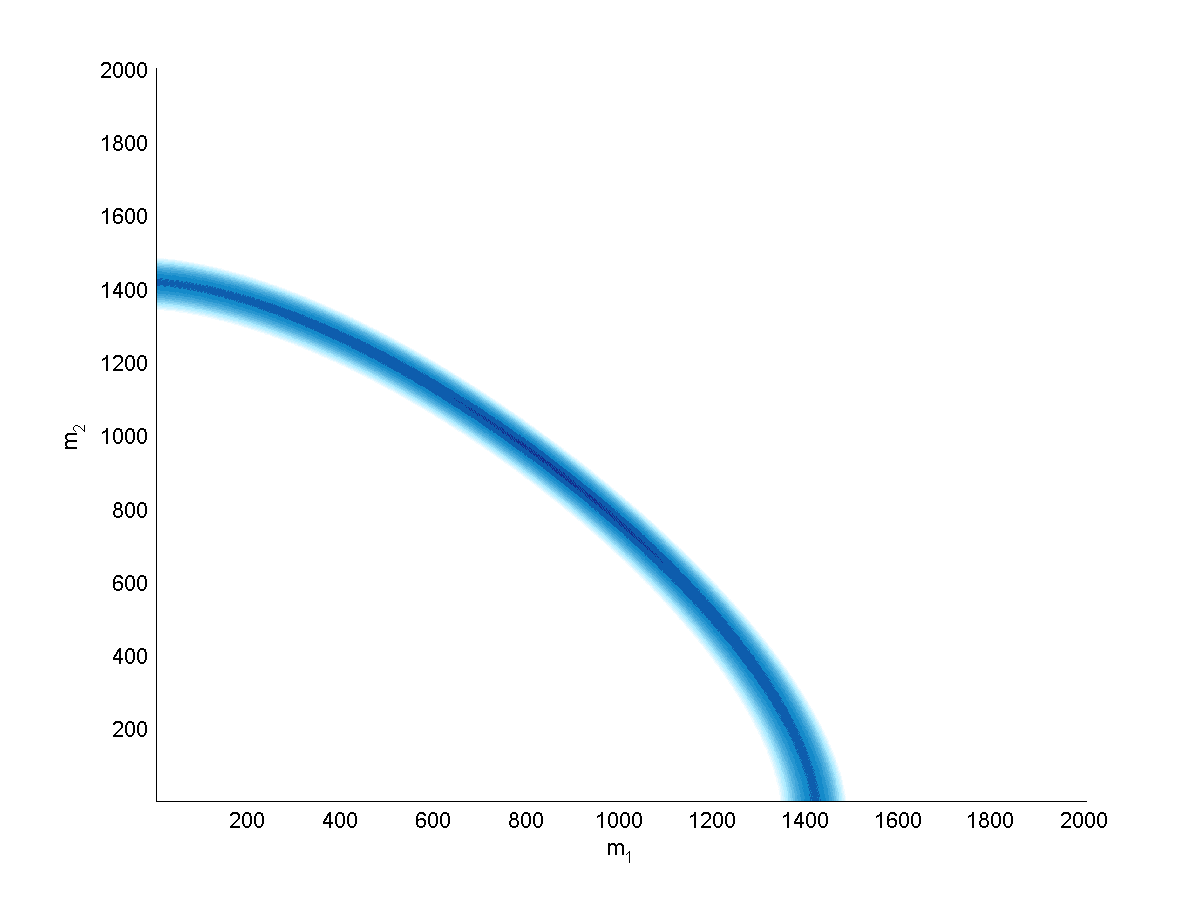}
						\caption{\label{fig:chi_b_c} \(\chi^2\) vs. the two endpoint masses for the \(\Lambda_c\) trajectory, for the two cases: \(\alp\) optimized (left) and \(\alp\) fixed at 0.95 GeV\(^{-2}\) (right). In the adjustable slope case the optimum has \(\alp = 1.2\) GeV\(^{-2}\). The red curves in the left plot are \(m_1 + m_2 = 2\times1000\) and \(m_1^{3/2}+m_2^{3/2}=2\times(1000)^{3/2}\). The real curve on which \(\chi^2\) is optimal, for which we do not know the exact parametrization, is between those two.}
	\end{figure}
	
	In the charmed baryon section we have only one trajectory we can use, comprised of three states, that of the \(\Lambda_c\) baryon. The best fits are again at a relatively high slope, 1.1 GeV\(^{-2}\), with the mass \(2m = 2010\) MeV. This fit's \(\chi^2\) tends to zero. The fit with the slope fixed at 0.95 GeV\(^{-2}\) takes the mass down to \(2m = 1760\) MeV with \(\chi^2 = 3\ten{-5}\). The high slope fit is equivalent to a fit with \(m_1 = 1720\) MeV and \(m_2 = 90\), while a the fit with \(m_1 = 1400\) and \(m_2 = 90\) is roughly equal to the latter fixed slope fit. We plot \(\chi^2\) as a function of the two endpoint masses for both the fixed and adjustable slope case in figure (\ref{fig:chi_b_c}).
	
	We can also do a fit using two \(\Xi_c\) states. These states are charmed and strange and are composed of \(dsc\) (\(\Xi_c^0\)) or \(usc\) (\(\Xi_c^+\)). Since we only have two states, we do only a fit with the fixed slope, \(\alp = 0.95\) GeV\(^{-2}\). The best massive fit then has \(2m = 2060\) MeV.

\section{Summary of results} \label{sec:summary}	
					\begin{table}[tp!] \centering
					\begin{tabular}{|c|c|cc|c|cc|} \hline
					Traj. & \(N\) & \multicolumn{2}{|c|}{\(m\)} & \alp & \multicolumn{2}{|c|}{\(a\)} \\ \hline
					
					\(N\) & \(7\) & \multicolumn{2}{|c|}{\(2m = 0-170\)} & \(0.944-0.959\) & \(a_e = (-0.32)-(-0.23)\) & \(a_o = (-0.75)-(-0.65)\) \\
					
					\(N\)\(^{[a]}\) & \(4\) & \multicolumn{2}{|c|}{\(2m = 0-640\)} & \(0.949-1.018\) & \(a_e = (-0.34)-0.50\) & \(a_o = (-0.98)-(-0.13)\) \\
					
					\(N\)\(^{[b]}\) & \(15\) & \multicolumn{2}{|c|}{\(2m = 0-425\)} & \(0.815-0.878\) & \(a_{1/2+} = (-0.22)-0.07\) & \(a_{3/2-} = (-0.36)-(-0.06)\) \\
					
					\(\Delta\) & \(7\) & \multicolumn{2}{|c|}{\(2m = 0-450\)} & \(0.898-0.969\) & \(a_e = 0.14-0.54\) & \(a_o = (-0.84)-(-0.42)\) \\
					
					\(\Delta\)\(^{[c]}\) & \(3\) & \multicolumn{2}{|c|}{\(2m = 0-175\)} & \(0.920-0.936\) & \multicolumn{2}{|c|}{\(a = 0.11-0.21\)} \\
					
					\(\Lambda\) & \(5\) & \multicolumn{2}{|c|}{\(2m = 0-125\)} & \(0.946-0.955\) & \multicolumn{2}{|c|}{\(a = (-0.68)-(-0.61)\)} \\
					
					\(\Sigma\) & \(3\) & \multicolumn{2}{|c|}{\(2m = 1190\)} & \(1.502\) & \multicolumn{2}{|c|}{\(a = (-0.15)\)} \\					 
					
					\(\Sigma\)\(^{[d]}\) & \(3\) & \multicolumn{2}{|c|}{\(2m = 1255\)} & \(1.459\) & \multicolumn{2}{|c|}{\(a = 1.37\)} \\
					
					\(\Xi\) & \(3\) & \multicolumn{2}{|c|}{\(2m = 1320\)} & \(1.455\) & \multicolumn{2}{|c|}{\(a = 0.50\)} \\
					
					\(\Lambda_c\) & \(3\) & \multicolumn{2}{|c|}{\(2m = 2010\)} & \(1.130\) & \multicolumn{2}{|c|}{\(a = 0.09\)} \\
					
					\hline \end{tabular}
					 \caption{\label{tab:summaryb} Summary table for the baryon fits. The ranges listed have \(\chi^2\) within 10\% of its optimal value.  \(N\) is the number of points in the trajectory. [a] is the (\(J,M^2\)) trajectory of the \(N\) baryons when taking only the four highest \(J\) states, [b] is a fit to radial trajectories of the \(N\). The fifteen states used are four states with \(J^P = \jph{1}{+}\), three with \jph{3}{-}, and four pairs with other values of \(J^P\). [c] is the radial trajectory of the \(\Delta\) (\jph{3}{+}). [d] is a trajectory beginning with the state \(\Sigma(1385)\) \jph{3}{+}, as opposed to the \jph{1}{+} \(\Sigma\) ground state. The rest of the trajectories are all leading trajectories in the \((J,M^2)\) plane, and do not exclude any states.}
					\end{table}

		\begin{table}[tp!] \centering
					\begin{tabular}{|c|c|cc|cc|} \hline
					Traj. & \(N\) & \multicolumn{2}{|c|}{\(m\)} & \multicolumn{2}{|c|}{\(a\)} \\ \hline
					
					\(N\) & \(7\) & \multicolumn{2}{|c|}{\(2m = 0-180\)} & \(a_e = (-0.33)-(-0.22)\) & \(a_o = (-0.77)-(-0.65)\) \\
								
					\(\Delta\) & \(7\) & \multicolumn{2}{|c|}{\(2m = 300-530\)} & \(a_e = 0.31-0.66\) & \(a_o = (-0.71)-(-0.26)\) \\
					
					\(\Lambda\) & \(5\) & \multicolumn{2}{|c|}{\(2m = 0-10\)} & \multicolumn{2}{|c|}{\(a = (-0.68)-(-0.61)\)} \\
					
					\(\Sigma\) & \(3\) & \multicolumn{2}{|c|}{\(2m = 530-690\)} & \multicolumn{2}{|c|}{\(a = (-0.29)-(-0.04)\)} \\
					
					\(\Sigma\)* & \(3\) & \multicolumn{2}{|c|}{\(2m = 435-570\)} & \multicolumn{2}{|c|}{\(a = 0.15-0.38\)} \\
					
					\(\Xi\) & \(3\) & \multicolumn{2}{|c|}{\(2m = 750-930\)} & \multicolumn{2}{|c|}{\(a = (-0.22)-0.10\)} \\
					
					\(\Lambda_c\) & \(3\) & \multicolumn{2}{|c|}{\(2m = 1760\)} & \multicolumn{2}{|c|}{\(a = (-0.36)\)} \\
					
					\(\Xi_c\) & \(2\) & \multicolumn{2}{|c|}{\(2m = 2060\)} & \multicolumn{2}{|c|}{\(a = (-1.13)\)} \\
					
					\hline \end{tabular}
					 \caption{\label{tab:fixedslope} \(J,M^2)\) fits done with the slope fixed at \(\alp = 0.950\) GeV\(^{-2}\). Fits with \(m_1 = m_2\) generally maximize \(m_1 + m_2\). In this table we may also include a fit for two \(\Xi_c\) states. The ranges listed have \(\chi^2\) within 10\% of its optimal value. \(N\) is the number of points in the trajectory.}
					\end{table}

We begin by presenting the two summary tables: in table (\ref{tab:summaryb}) we list the results of the general fits, while in table (\ref{tab:fixedslope}) are the results of the fits with the fixed slope, \(\alp = 0.95\) GeV\(^{-2}\).

The internal structure of the baryon is more complex than that of the meson, and as a result a unified stringy model of the baryon is harder to construct. Out of the various stringy models we have examined, the model of a quark and diquark is best supported by experiment. This is based mostly on the observation that the Regge slope for baryons is roughly equal to the meson slope - so if we assume a universal string tension, the baryons must be described by a single string model as the mesons are. As explained in section \ref{sec:barsym}, we present our results in terms of the total mass of the endpoints, as we cannot determine the distribution of the mass between them from the Regge trajectory fits alone.

Our massive fits for the baryons are not always consistent. For the light quark trajectories, of the \(N\) and the \(\Delta\), we have seen there is no evidence for a string endpoint mass of the light quarks. These states are best fitted by linear trajectories with a slope similar to that of the light mesons - around 0.95 GeV\(^{-2}\) for the \(N\) and \(0.90\) GeV\(^{-2}\) for the \(\Delta\). The \((J,M^2)\) trajectories also exhibit a splitting between the trajectories of states with even and odd orbital angular momentum. In our fits we incorporate this difference into the intercept alone.

Since we do not see the splitting effect in the trajectory of the strange \(\Lambda\) baryon, which has five data points, we assume this effect is only significant for the light baryons. The next heavier trajectories after the \(\Lambda\) are comprised of only three data points and therefore cannot be used to determine whether this is a correct assumption or not. In the following we summarize our results using this assumption.

For the strange baryon trajectories there is a significant discrepancy between the \(\Lambda\), which is best (and very well) fitted by a linear, massless, trajectory, and the \(\Sigma\) baryons which are optimally fitted with a high total mass, of around 1200 MeV and an unusually high slope, 1.5 GeV\(^-2\). The fits when fixing the slope at the value obtained from the lighter baryon fits give a total mass in the more reasonable \(500-600\) MeV range. We have also looked into the doubly strange \(\Xi\) baryon, where we have a similar picture: the same high slope (1.5) and masses (1300) or \(\alp = 0.95\) GeV\(^{-2}\) and a total mass of around \(800\) MeV. We noted that the latter choice, of a fixed slope, is not only more consistent with the slopes obtained from the lighter baryons, but also with the mass obtained for the \(s\) quark in the meson fits, \(m_s \approx 400\) MeV.

The last results are those of the charmed baryons, the heaviest baryons for which we have a trajectory.  The charmed \(\Lambda_c\) is again best fitted by a high slope, 1.2GeV\(^{-2}\) and a total mass of a little over 1800 MeV. Once more we can bring down the mass by fixing \(\alp\) at 0.95 GeV\(^{-2}\) and get \(2m = 1760\) MeV as the best fit. For the charmed-strange \(\Xi_c\) trajectory, including only two data points and therefore fitted only with the fixed slope of 0.95 GeV\(^{-2}\), we find a fit with \(2m = 2060\) MeV.

\subsection{Comparison with meson fits}
This section will briefly summarize the results of the fits to the Regge trajectories of mesons done in \cite{Sonnenschein:2014jwa}, and offer a comparison between them and the results presented in this paper for the baryon trajectories.

The mesons are expected to have a simpler structure than the baryons. Therefore, one could say they are a better source from which we can begin to extract the parameters of our models. We have only one stringy model of the meson - in holography it is the rotating open string connected at both ends to flavor branes, and in the mapping to flat space-time it is a single string with two endpoint masses.

The first result, that is easiest to compare, is that of the Regge slope. For the \((J,M^2)\) trajectories of the mesons we found that, with added masses, all trajectories involving \(u\), \(d\), \(s\), and \(c\) quarks are well fitted with a Regge slope of around \(0.9\) GeV\(^{-2}\).\footnote{The only exception, in the \((J,M^2)\) plane, was the trajectory of the \(\Upsilon\) (\(b\bar{b}\)), which had a slope of 0.64 GeV\(^{-2}\). For the baryons, we do not have trajectories of bottom baryons, and therefore, no point of comparison. In \cite{Sonnenschein:2014jwa} we noted that since our model is based on a long string approximation it might not hold for the heavy \(b\bar{b}\) mesons, and hence the discrepancy in the slope. We do not have trajectories of baryons containing \(b\) quarks, and the problem of short strings is less apparent in the current analysis.} For the baryons, we have seen that the light (\(N\) and \(\Delta\)) baryons, as well as the strange \(\Lambda\), are best fitted with a slope of 0.90-0.95 GeV\(^{-2}\). This similarity between the meson and baryon results is what drove us to use the single string quark-diquark model for the baryon, which predicts an equal Regge slope between baryons and mesons.

Moving on to the heavier baryons, we found higher slopes (of up to 1.5 GeV\(^{-2}\)) but also found that we can get reasonably good fits using 0.95 GeV\(^{-2}\) as a ``universal'' slope, common to all available trajectories (light, strange, and charmed). We have seen this happen for the mesons as well, but the differences there between the optimal fits and the ``universal slope'' fits were smaller - there it was the difference between 1.0-1.1 and 0.9 GeV\(^{-2}\).

In the \((n,M^2)\) plane the light meson trajectories had a similar slope to their respective orbital trajectories, but always slightly lower. A generic value obtained there would be 0.8 GeV\(^{-2}\). For the baryons, we only analyzed the trajectories of the \(N\) and the \(\Delta\). For the \(N\) we have seen the same behavior we have for the mesons, with approximately the same slope, but for the \(\Delta\) the slope in \((n,M^2)\) is the higher one.

Knowing the different mesons' compositions, we can extract from them the quark masses directly. The two endpoint masses correspond directly to the quark and anti-quark making up the meson, and so we naturally identify the masses obtained in our fits with quark masses.

The mass of the \(u\) and \(d\) quarks we could not determine from the meson fits. We found no evidence that clearly states that the light quarks have a non-zero mass, but we did not exclude a mass of up to about 100 MeV. For the \(s\) quark meson fits we found that a non-zero mass for the \(s\) was always preferable. The results were generally in the range \(m_s = 200-400\) MeV. The \(c\) quark was found to have a mass close to the value of its mass as a constituent quark, around 1500 MeV.

The light baryon trajectories seem to prefer massless endpoints without completely excluding masses of a few dozen MeV, or up to a hundred MeV. This is similar to the meson result. The strange \(\Lambda\) offers the biggest discrepancy in terms of the mass. The \(\Lambda\) trajectory is best fitted by a simple linear fit with no endpoint masses where we would have expected the presence of the \(s\) quark to contribute a mass of at least 200 MeV. The other strange baryons, on the other hand, seem to be consistent with the meson results, especially when fixing the slope at its universal value. The results for the charmed \(\Lambda_c\) baryon are also consistent with a mass of 1500 MeV for the \(c\) quark, as is the result for the charmed-strange baryon, which is compatible with a mass of 1500 MeV on one end (for the charmed), and a mass of 400 MeV on the other.

%\subsection{The intercept}
%In previous sections we have only discussed the results for the slope and the endpoint masses. The behavior of the remaining fitting parameter - the intercept - is also interesting and should be studied.

\subsection{Structure of the baryon in the quark diquark model} \label{sec:structure}

For the light baryons our analysis of the spectrum cannot offer much new insight regarding the different baryons' structure,\footnote{\cite{Selem:2006nd} offers a discussion of the composition of the light baryons in a model of a quark and diquark joined by a flux tube. In the analysis of the spectrum done there, the light baryons are assigned different configurations of the diquark based on the energetics of the \(ud\) diquarks.} in particular because we have no way to distinguish between the two light quarks, given their small - possibly zero - masses, but also because we cannot in general make any comments regarding the mass distribution within the different baryons (both light and heavy). In spite of this, there is one interesting implication when interpreting our results in light of the underlying holographic models and the way they map the diquarks into flat space-time.

\begin{figure}[t!] \centering
					\includegraphics[width=.90\textwidth, natheight=1056bp,natwidth=1902bp]{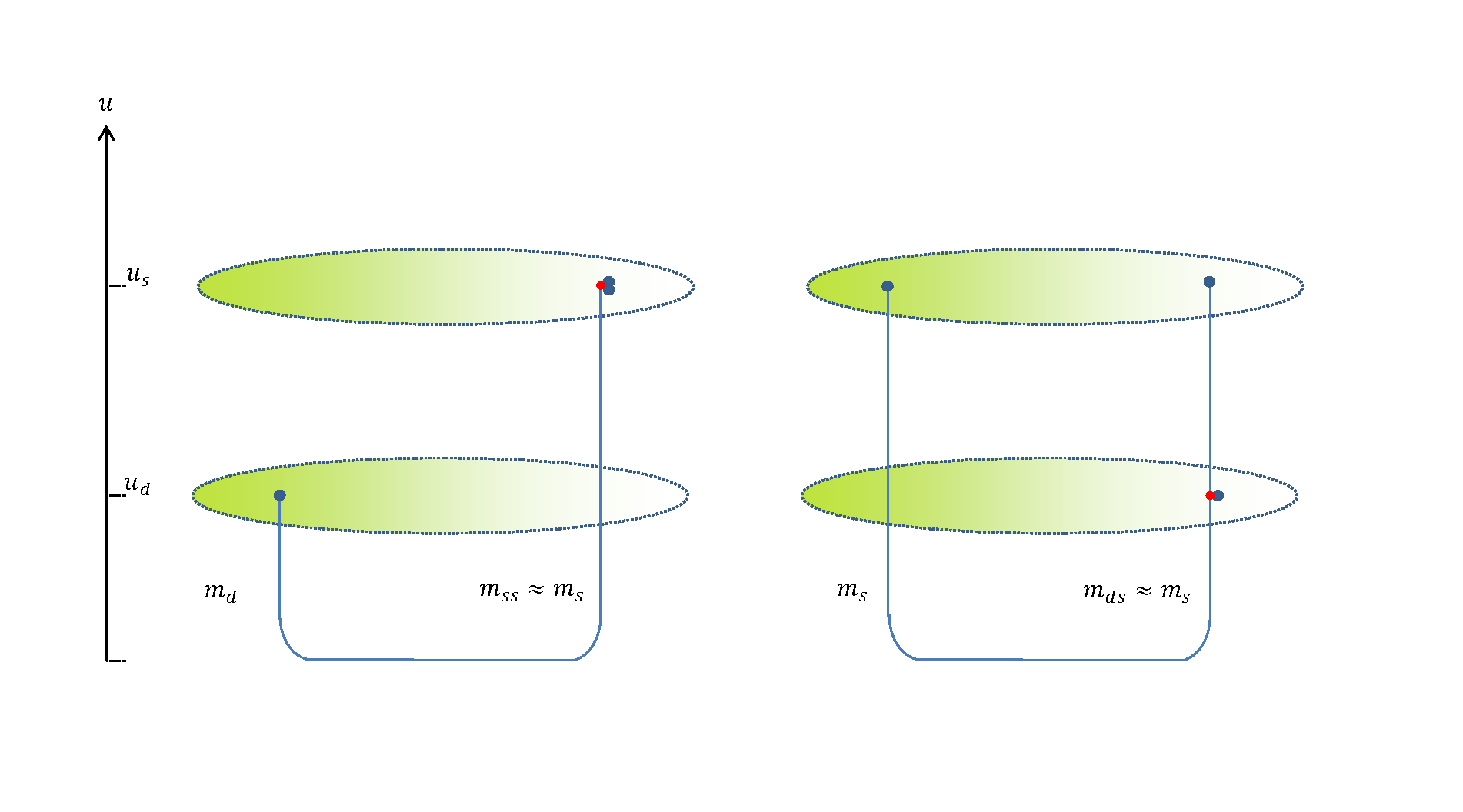}
					\caption{\label{fig:ssdbaryon}  The two holographic setups for the \(\Xi^-\) baryon, with different compositions of the diquarks. The vertical segments of the strings contribute the measured endpoint masses.}
\end{figure}

In our analysis of the meson spectrum, we argued that the mass parameter relevant to the analysis is the mass of the quark as a string endpoint, which generically was found to be between the usual QCD and constituent masses attributed to the respective quark. For the diquark the identification between string length and mass can have another implication, as illustrated in figure (\ref{fig:ssdbaryon}). If the relevant mass parameter is the length of the vertical segment of the string connected to the flavor brane, and if the two quarks forming the diquark and the baryonic vertex to which they are both connected all lie close to each other on the flavor brane, then we would expect the mass of the diquark - in the holographic picture, as a string endpoint - to be approximately equal to the mass of a single quark:

\be m_{qq} \approx m_q \ee

This is because we have only one contribution to the mass from the string connecting the lone quark outside the diquark and the baryonic vertex. This is a prediction that can serve as a test of the holographic interpretation of the string endpoint masses. Since we do not have an accurate figure for the mass of the light \(u\) and \(d\) quarks, and since we lack data for charmed and heavier baryons, our best avenue for verifying this experimentally is by examining the doubly strange \(\Xi\) baryon.

The two options for the \(\Xi\) quark are one where the diquark is composed of an \(s\) and a light quark, and another where the diquark is composed of two \(s\) quarks. For the first option, our holographic interpretation would lead us to expect there to be two masses at the endpoints approximately equal to the \(s\) quark mass, leading to a total mass of \(2m \approx 2m_s\) at the endpoints. In the second option, the two \(s\) quarks in the diquark would contribute only once to the total mass we measure in the Regge trajectory fits, so the result for the total mass \(2m\) should be around, possibly a little higher than, the mass of a single \(s\) quark.

\begin{figure}[t!] \centering
					\includegraphics[width=.90\textwidth, natheight=1122bp,natwidth=1944bp]{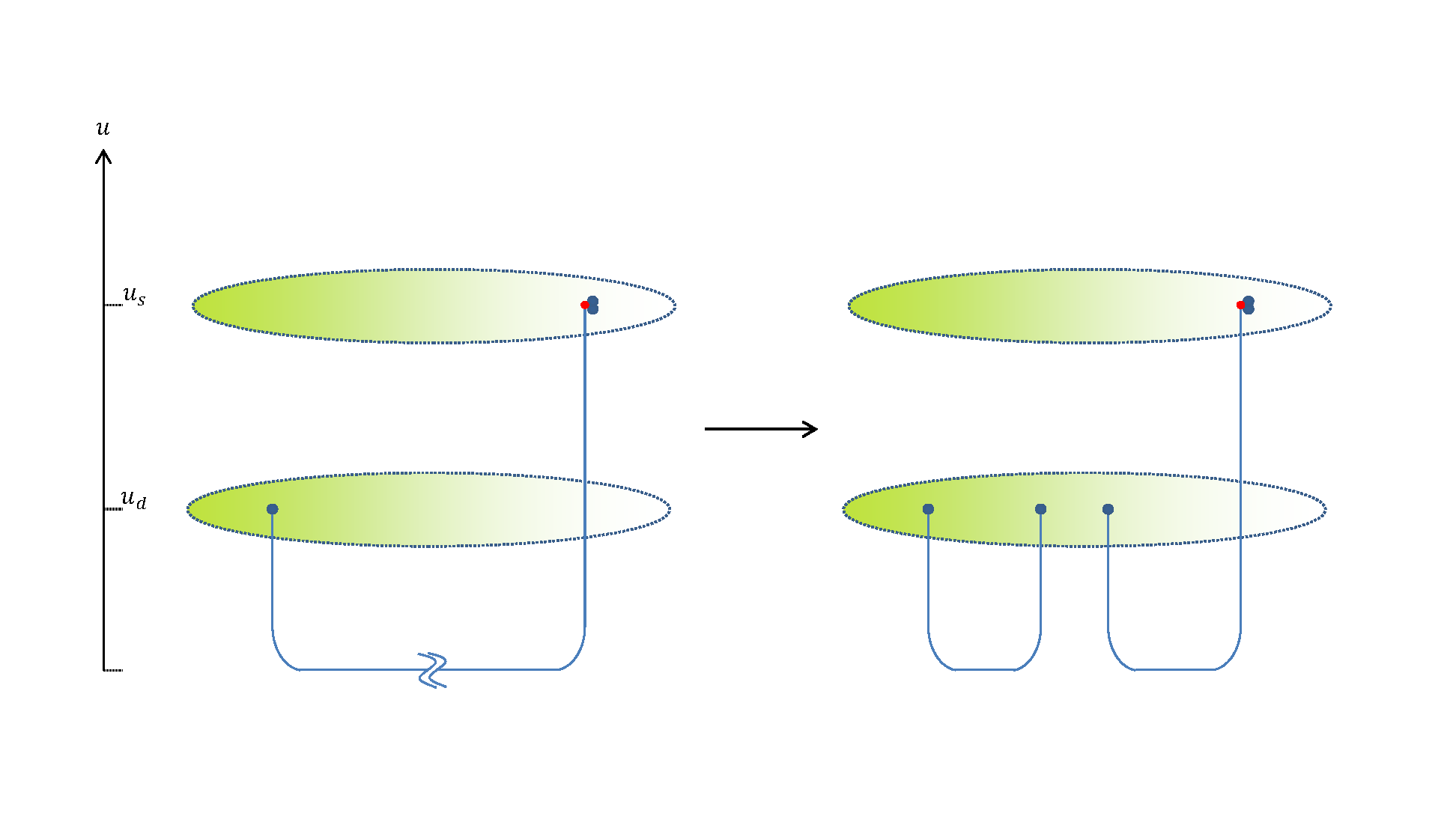}
					\caption{\label{fig:ssddecay}  A doubly strange baryon with an \(ss\) diquark decaying into a doubly strange baryon and a non-strange meson. First the string tears, and then the endpoints reconnect to the flavor brane, forming a quark-anti-quark pair.}
\end{figure}

The result from the fixed slope fit of the \(\Xi\) trajectory, \(2m = 750-930\) MeV, is consistent, from the holographic point of view, with a \(ds\) or \(us\) diquark, as opposed to \(ss\). This is because we expect the mass of the \(s\) to be somewhere near \(400\) MeV. Of course, from a purely flat space-time perspective, an \(ss\) diquark with a mass of roughly \(2m_s\) is not excluded.

If we look at the decay modes of the states used in the \(\Xi\) fits we might learn something about their structure \cite{Peeters:2005fq}\cite{Decays}. We look at a baryon's strong decays into a baryon and a meson, and our assumption is that in these decays the diquark and baryonic vertex go into the outgoing baryon while the third lone quark ends up in the meson. An illustration of this type of decay is in figure (\ref{fig:ssddecay}).

The lightest doubly strange state does not have the phase space for strong decays. If we look at the two next states in the trajectory, the \(\Xi(1820)\) and the \(\Xi(2030)\), we see that they decay mainly into \(\Lambda K\) or \(\Sigma K\). The fact that they decay into a strange meson and strange baryon is against the \(ss\) diquark configuration. The leading modes of decay should leave the diquark intact, so if the diquark were \(ss\) the leading mode of decay would be \(\Xi\pi\) (as it is for some of the other observed doubly strange baryons).

For the baryons with a single strange quark, the \(\Lambda\) and the \(\Sigma\), we have seen an odd discrepancy between the obtained mass values. The mass in the \(\Lambda\) baryon was less than \(100\) MeV, while in the \(\Sigma\) we have seen masses of above \(400\) MeV. We cannot explain this discrepancy in terms of different configurations of the diquarks, as we expect the \(s\) to contribute to the mass whether it is in the diquark or not. The decay modes do not give a straightforward answer regarding the compositions of the \(\Lambda\) and \(\Sigma\), as the states decay both to \(NK\) and \(\Sigma \pi\)/\(\Lambda\pi\). We do not see a systematic preference for decays where the \(s\) remains in the baryon (implying it is near the baryonic vertex in the diquark) or vice versa in either of the trajectories.

For the charmed-strange \(\Xi_c\) we see a mass compatible compatible with \(m_1 + m_2 = m_s + m_c\). This implies to us that the possibility of a \(cs\) diquark is excluded, since we see both quarks' masses (from the holographic point of view we expect the diquark mass to be \(m_{cs} \approx m_c\)). Unfortunately we cannot test this based on the decay modes. If we look at the decays of the \(\Xi_c\) baryons, we find that the \(\Xi_c^0\)/\(\Xi_c^-\) does not have the phase space to decay strongly, and the next state we take in the trajectory, \(\Xi_c(2815)\), is also too light to provide information that would be useful to us. The \(\Xi_c(2815)\) cannot decay to a charmed meson and a strange baryon (which is the decay mode we will naively expect if the \(\Xi_c\) is a \(us\)/\(ds\) diquark joined to a \(c\) quark), simply because the lightest of these, \(D^\pm\)/\(D^0\) and \(\Lambda\) respectively, are still heavy enough so that the sum of their masses exceeds the mass of the \(\Xi_c(2815)\).

For the charmed \(\Lambda_c\) baryon, with one \(c\) and two light \(u/d\) quarks, we have no prediction based on the masses, because in any case we expect to see a total mass of approximately \(m_c = 1500\) MeV. The first state in the trajectory heavy enough to decay into a charmed meson and a baryon, the \(\Lambda_c(2800)^+\), was observed to decay both to \(pD^0\) and \(\Sigma_c \pi\), but there is no quantitative data to indicate which of these modes (if any) is dominant.

\section{Conclusions and future directions} \label{sec:conclusions}
If we seek a universal stringy model of the baryon the best option seems to be the model of a quark and a diquark at the endpoints of a rotating string. We have seen how this model can be obtained from a mapping of rotating holographic strings to flat space-time, and it is the only stringy configuration in flat space-time which we know to be classically stable.

From a phenomenological point of view, unlike the three string Y-shape or the closed string \(\Delta\)-shape models, the quark-diquark model is supported by the simple fact that the baryons lie on Regge trajectories with a slope roughly equal to that of the meson trajectories. In our fitting analysis we have presented fits of trajectories of baryons composed of \(u\), \(d\), \(s\), and \(c\) quarks, an in particular fits with the fixed slope \(\alp = 0.95\) GeV\(^{-2}\). The fixed slope fits, while not optimal, are for the most part consistent with the meson fit results of \cite{Sonnenschein:2014jwa}.

In section \ref{sec:structure} we have explained how one could test the holographic interpretation of the endpoint masses as the lengths of the vertical segments of the string along the radial dimension by examining the trajectories of doubly heavy baryons. We have attempted this test for the doubly strange \(\Xi\) baryon, but found no conclusive answer. In that section we also briefly discussed our approach to the holographic decay of hadrons. A detailed analysis of hadronic decays, not only qualitative as offered in this paper but also quantitative, is one avenue for continuing the use of our model for research. The static properties of baryons have also been examined in holography \cite{Hata:2007mb}\cite{Seki:2008mu}, and it is left to see how one can describe some of them using a stringy model.

Another prediction from holography is the presence of a baryonic vertex. Our fits exclude the presence of a baryonic vertex mass at the center of mass of the rotating baryon, but in the quark-diquark model which we prefer it is expected to be found at one of the string endpoints, with the diquark. If there is a baryonic vertex at an endpoint, there is no evidence to suggest it contributes greatly to the endpoint mass.

We could also attempt enhancements of the model. Adding spin degrees of freedom to the endpoints would give us a much better chance of constructing a universal model that would describe the entire baryon spectrum. In \cite{Selem:2006nd} the distinction between spin zero and spin one diquarks played an important part in analyzing the spectrum, while we have only discussed the flavor structure of the diquark and ascribed its mass to the holographic string alone (with a possible small addition from the baryonic vertex). Spin interactions could also help explain the even-odd splitting observed in the light baryons - our simple classical model will not explain a phenomenon that distinguishes between symmetric and anti-symmetric states without some additional interaction.

We should also strive to gain a better understanding of the intercept. While all previous sections discuss results for the slope and endpoint masses alone, the intercept, the results for which are listed in the summary tables of section (\ref{sec:summary}), is an interesting parameter from a theoretical point of view, and understanding its behavior is an important goal in constructing a truly universal model of the baryon. Added interactions should contribute, in leading order, a correction to the intercept, which should also be affected by the endpoint masses, and understanding the intercept's behavior may also help us distinguish between different configurations of the baryon without requiring additional information from experiment.

\clearpage
\appendix

\section{Individual trajectory fits} \label{app:individualb}

This section presents the data and individual trajectory fits. The experimental data is taken from the Particle Data Group's (PDG) Review of Particle Physics \cite{PDG:2012}.
	
\subsection{The states used in the fits}

\begin{table}[t!] \centering
	\begin{tabular}{|c|c|c|l|l|c|c|c|l|l|} \hline
		Traj. & \(I\) & \(J^{P}\) & State & Status & Traj. & \(I\) & \(J^{P}\) & State & Status\\ \hline
		
		\(N\) & \(\frac{1}{2}\)
						& \jph{1}{+}  & \(n/p\)     & **** 		&	 \(\Lambda\) & \(0\)
																									& 	\jph{1}{+}  & \(\Lambda(1116)\) & **** 	\\
					& & \jph{3}{-}  & \(N(1520)\) & **** 	&	\(S = -1\) & & \jph{3}{-}  & \(\Lambda(1520)\) & **** 	\\
					& & \jph{5}{+}  & \(N(1680)\) & ****	& & & \jph{5}{+}  & \(\Lambda(1820)\) & ****	\\
					& & \jph{7}{-}  & \(N(2190)\) & ****	& & & \jph{7}{-}  & \(\Lambda(2100)\) & ****	\\
					& & \jph{9}{+}  & \(N(2220)\) & ****	& & & \jph{9}{+}  & \(\Lambda(2350)\) & ***		\\ \cline{6-10}
					& & \jph{11}{-} & \(N(2600)\) & *** 		&	\(\Sigma\) & \(1\)
																											& 	\jph{1}{+} & \(\Sigma(1193)\) & **** \\
					& & \jph{13}{+} & \(N(2700)\) & ** 				&	\(S = -1\) & & \jph{3}{-} & \(\Sigma(1670)\) & **** \\ \cline{1-5}
		\(\Delta\) & \(\frac{3}{2}\)
						& \jph{3}{+}  & \(\Delta(1232)\) & **** & & & \jph{5}{+} & \(\Sigma(1915)\) & **** \\ \cline{6-10}
					& & \jph{5}{-}  & \(\Delta(1930)\) & *** 		& \(\Sigma\) & \(1\)
																											&		\jph{3}{+} & \(\Sigma(1385)\) & **** \\
					& & \jph{7}{+}  & \(\Delta(1950)\) & ****	& \(S = -1\) & & \jph{5}{-} & \(\Sigma(1775)\) & **** \\	
					& & \jph{9}{-}  & \(\Delta(2400)\) & **		& & & \jph{7}{+} & \(\Sigma(2030)\) & **** \\ \cline{6-10}
					& & \jph{11}{+} & \(\Delta(2420)\) & ****	& \(\Xi\) & \(\frac{1}{2}\)
																										& 	\jph{1}{+} & \(\Xi^0/\Xi^-\)  & ****\\
					& & \jph{13}{-} & \(\Delta(2750)\) & **	& \(S = -2\) & & \jph{3}{-} & \(\Xi(1820)\) & ***\\
					& & \jph{15}{+} & \(\Delta(2950)\) & **	& & & \jph{5}{+} & \(\Xi(2030)\) & *** \\ \cline{1-5}\cline{6-10}
					& & &	&																		& \(\Lambda_c\) & \(0\)
																										& 	\jph{1}{+} & \(\Lambda_c(2286)^+\) & **** \\
					& & &	&																	& \(C = 1\) & & \jph{3}{-} & \(\Lambda_c(2625)^+\) & *** \\
					& & & &																	& & & \jph{5}{+} & \(\Lambda_c(2880)^+\) & *** \\ \cline{6-10}
					
										& & &	&																		& \(\Xi_c\) & \(\frac{1}{2}\)
																										& 	\jph{1}{+} & \(\Xi_c^+/\Xi_c^0\)) & *** \\
					& & &	&																	& \(C = 1\) & & \jph{3}{-} & \(\Xi_c(2815)^+\) & *** \\
					& & & &                                 & \(S = -1\) & & & & \\ \hline
					
	\end{tabular}
	\caption{\label{tab:baryons_j} The baryon states used in the \((J,M^2)\) trajectory fits.}
\end{table}

\begin{table}[t!] \centering
	\begin{tabular}{|c|c|c|l|l|c|c|c|l|l|} \hline
		Traj. & \(I(J^P)\) & \(n\) & State & Status & Traj. & \(I(J^P)\) & \(n\) & State & Status \\ \hline
		\(N\) & \(\frac{1}{2}(\jph{1}{+})\) & \(0\) & \(n/p\) & **** & \(N\) & \(\frac{1}{2}(\jph{1}{-})\) & 0 & \(N(1535)\) & **** \\
		 & & \(1\) & \(N(1440)\) & ****	& & & 1 & \(N(1895)\) & ** \\ \cline{6-10}
		 & & \(2\) & \(N(1880)\) & **  & \(N\) & \(\frac{1}{2}(\jph{3}{+})\) & 0 & \(N(1720)\) & **** \\
		 & & \(3\) & \(N(2100)\) & *   & & & 1 & \(N(2040)\) & * \\ \cline{1-5} \cline{6-10}
		\(N\) & \(\frac{1}{2}(\jph{3}{-})\) & \(0\) & \(N(1520)\) & **** & \(N\) & \(\frac{1}{2}(\jph{5}{-})\) & 0 & \(N(1675)\) & **** \\
		 & & \(1\) & \(N(1875)\) & *** & & & 1 & \(N(2060)\) & ** \\ \cline{6-10}
		 & & \(2\) & \(N(2150)\) & **  & \(\Delta\) & \(\frac{3}{2}(\jph{1}{+})\) & 0 & \(\Delta(1232)\) & ****\\ \cline{1-5}
		\(N\) & \(\frac{1}{2}(\jph{5}{+})\) & \(0\) & \(N(1680)\) & **** & & & 1 & \(\Delta(1600)\) & *** \\
		 & & \(1\) & \(N(2000)\) & ** & & & 2 & \(\Delta(1920)\) & *** \\ \hline
	\end{tabular}
	\caption{\label{tab:baryons_n} The baryon states used in the \((n,M^2)\) trajectory fits and their assignments.}
\end{table}

The states we have used in our analysis, and their assignment into trajectories, are summarized in tables (\ref{tab:baryons_j}) and (\ref{tab:baryons_n}). The first table is for the \((J,M^2)\) plane and the latter for the \((n,M^2)\) trajectories. For the baryons, we also indicate the overall ``status'' of their PDG listing, ranking the resonances from one to four stars based on how well established are they and their properties in experiment. As a general rule we consider states with three or four stars safe to use as we see fit. We include lesser resonances only when they complement a trajectory formed by well established states.

To build the trajectories, we assume the relation between a states orbital angular momentum and its parity is \(P = (-1)^L\), so even and odd states in a given trajectory in the \((J,M^2)\) have alternating parity, as they do for mesons. We do fits only to leading trajectories (with \(n = 0\)), so we always select the lightest mass state for a given angular momentum and with the appropriate quantum numbers for a given trajectory.

The majority of states used in the \((J,M^2)\) fits are states with a PDG status of four or three stars. The few exceptions are the high spin states of the \(N\) and \(\Delta\) baryons, where we use some two star states. These states generally fit in well with their respective trajectories, which might be considered as evidence for the existence of those high \(J\) states at their given masses. The only exception is the \(\Delta(2950)\) which lies above its predicted place in the trajectory. Its mass should be about 100 MeV lower than the mass the PDG lists for it.

The strange baryons used are all well established states, but there one comment to be made about the \(\Sigma\) states chosen. In the first trajectory, beginning with the \(J^P = \jph{1}{+}\) \(\Sigma\) ground state, the next state chosen (with \jph{3}{-}) is the \(\Sigma(1670)\). The PDG lists another state with the same \(J^P\) at a lower mass, \(\Sigma(1580)\), but it is a one star state and its existence is very uncertain. If we use the \(\Sigma(1580)\) as the \jph{3}{-} state in the trajectory, we see that it is well fitted by a linear trajectory with \(\alp = 0.90\) GeV\(^{-2}\). The choice of \(\Sigma(1670)\) is more compatible with the massive fit, and has similar results to the ones obtained from the second \(\Sigma\) trajectory, where we use the lightest known states with \(J^P = \jph{3}{+}\), \jph{5}{-}, and \jph{7}{+}, all well established.

For the doubly strange trajectories we use, alongside the ground state \(\Xi^0/\Xi^\pm\) and the \(\Xi(1820)\), the state \(\Xi(2030)\). This is a three star state in the PDG, but its parity and angular momentum are not exactly known. The PDG places the bound \(J \geq 5/2\) on its angular momentum, and our analysis of the Regge trajectory seems consistent when identifying this state as the \(J^P = \jph{5}{-}\) state in the leading \(\Xi\) trajectory.

Going to the charmed baryon sector, we still have well established states we can use, but some of these states' quantum numbers are yet to be measured directly. The second state in the \(\Lambda_c\) trajectory is the \(\Lambda_c(2625)\), for which \(J\) and \(P\) have not been measured, but the PDG offers the assignment \(J^P = \jph{3}{-}\). In the charmed-strange sector, the two states \(\Xi_c\) states we use are again without a direct measurement of their spin-parity, and are only assigned \(J^P = \jph{1}{+}\) and \jph{3}{-}. Our analysis of the Regge trajectories gives us no reason to doubt these assignments.

We also include some trajectories of the light baryons in the \((n,M^2)\) plane. For the \(\Delta\) baryons we have two states with three stars following the well known and well established ground state. For the \(N\) baryons, we present an assignment of fifteen states into six trajectories, each with a different spin-parity. Here is the only place where we use some dubious one star states. We include them mostly to observe how this assignment results in parallel trajectories in the \((n,M^2)\) plane. The conclusions we eventually draw from the fits pertain to the value of the slope obtained, and the slope does not change significantly if we include only the best established states, the \(J^P = \jph{1}{+}\) or \(\jph{3}{-}\) \(N\) baryons with three or four PDG stars. \(N\) resonances with three or four stars that we do not use in these fits are the \(N(1650) \jph{1}{-}\), \(N(1700)\) \jph{3}{-}, \(N(1710)\) \jph{1}{+}, and \(N(1900)\) \jph{3}{+}. For each of these we find at least one lower resonance with the same \(J^P\), and all are too low in mass to be the next states in their respective trajectories.

\subsection{Trajectories in the \texorpdfstring{$(J,M^2)$}{(J,M2)} plane}
			
			\subsubsection{Light quark baryons}
			
			\begin{figure}[t!] \centering
						\includegraphics[natwidth=1200bp, natheight=900bp, width=.48\textwidth]{bar_j_n.png}	\hfill
						\includegraphics[natwidth=1200bp, natheight=900bp, width=.48\textwidth]{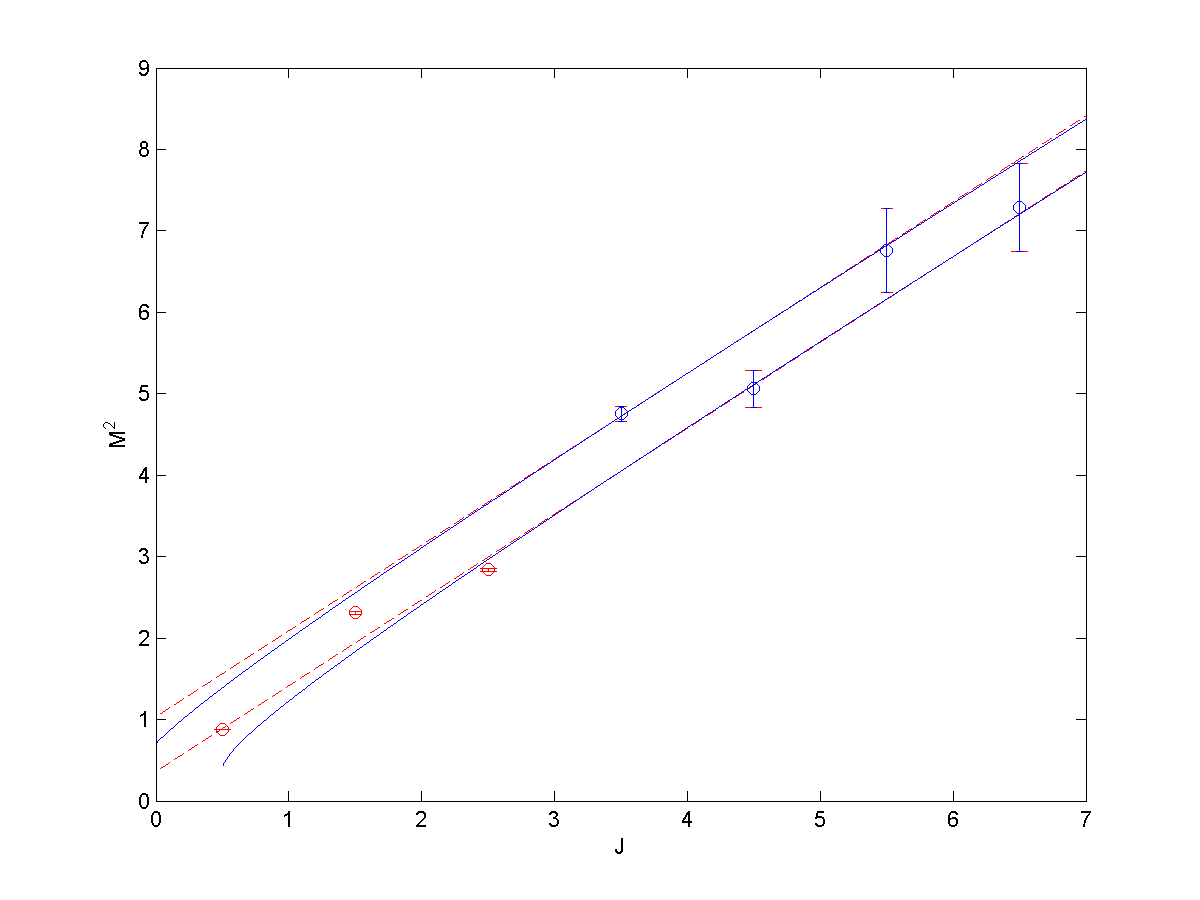} \\
						\includegraphics[natwidth=1200bp, natheight=900bp, width=.48\textwidth]{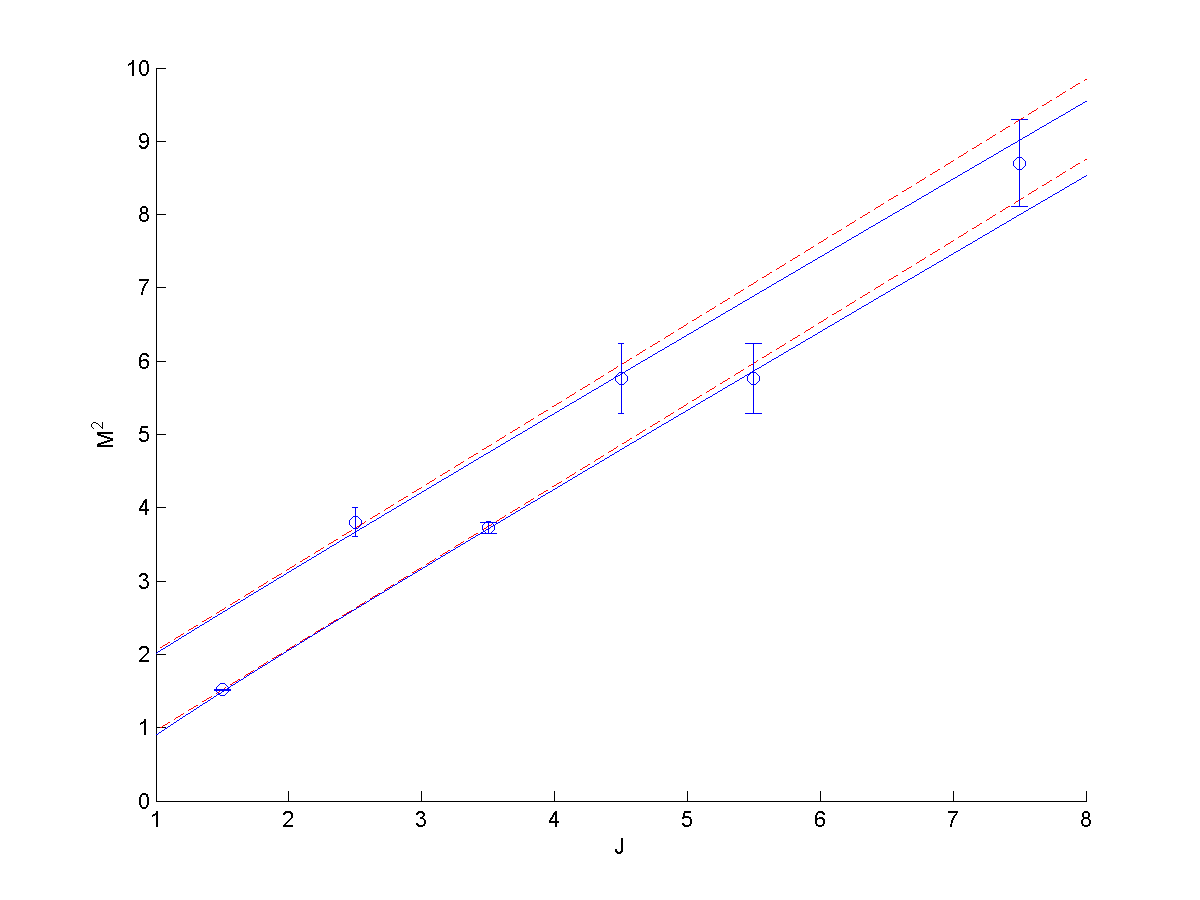}
						\caption{\label{fig:bar_j_light} Top: The \(N\) with linear (red, dashed line) and maximal mass fits (blue line). The maximal mass is defined as the highest mass with \(\chi^2\) within 10\% of its optimal value. Top-left: all the states and fit with \(2m = 170\). Top-right: fit to highest \(J\) states alone leads to a maximal mass of \(2m = 640\). The states marked in red are those excluded from the fits. Bottom: Trajectory of the \(\Delta\) with the maximal mass \(2m = 450\).}
				\end{figure}
			
			The states in this section are all comprised of \(u\) and \(d\) quarks only.
				
				\paragraph{The \(N\) trajectory:} The states, with their \(J^P\) values, are \(N(939) \frac{1}{2}^+\), \(N(1520) \frac{3}{2}^-\), \(N(1680) \frac{5}{2}^+\), \(N(2190) \frac{7}{2}^-\), \(N(2220) \frac{9}{2}^+\), \(N(2600) \frac{11}{2}^-\), and \(N(2700) \frac{13}{2}^+\). The trajectory and its fits are depicted in (\ref{fig:bar_j_light}).
				
				The linear fit is
				\[ \alp = 0.944, a_e = -0.32, a_o = -0.75 \]
				with \(\chi^2_l = 12.15\ten{-4}\). It is optimal, and the highest mass fit with a good \(\chi^2\) is
				\[ 2m = 170, \alp = 0.959, a_e = -0.23, a_o = -0.65\]
				with \rchi{1.10}.
				
				When taking only the four highest \(J\) states in the trajectory (i.e. the states starting from \(J = 7/2\)), the mass dependence is weaker. The linear fit, now
				\[ \alp = 0.949, a_e = -0.34, a_o = -0.98 \]
				is still optimal, with \(\chi^2_l = 1.10\ten{-4}\), but we can go to higher masses, such as
				\[ 2m = 640, \alp = 1.018, a_e = 0.50, a_o = -0.13 \]
				which has \rchi{1.09}.
				
				\paragraph{The \(\Delta\) trajectory:} Here we use the states \(\Delta(1232)\frac{3}{2}^+\) , \(\Delta(1930) \frac{5}{2}^-\), \(\Delta(1950) \frac{7}{2}^+\), \(\Delta(2400)\frac{9}{2}^-\), \(\Delta(2420)\frac{11}{2}^+\), \(\Delta(2750)\frac{13}{2}^-\), and \(\Delta(2950)\frac{15}{2}^+\). The linear fit is again optimal
				\[ \alp = 0.898, a_e = 0.14, a_o = -0.84 \]
				and it has \(\chi^2_l = 12.51\ten{-4}\). The highest mass fit is
				\[ 2m = 450, \alp = 0.969, a_e = 0.54, a_o = -0.42 \]
				with \rchi{1.10}. This trajectory is depicted in the bottom plot of figure (\ref{fig:bar_j_light}).
				
				\begin{figure}[t!] \centering
						\includegraphics[natwidth=1200bp, natheight=900bp, width=.48\textwidth]{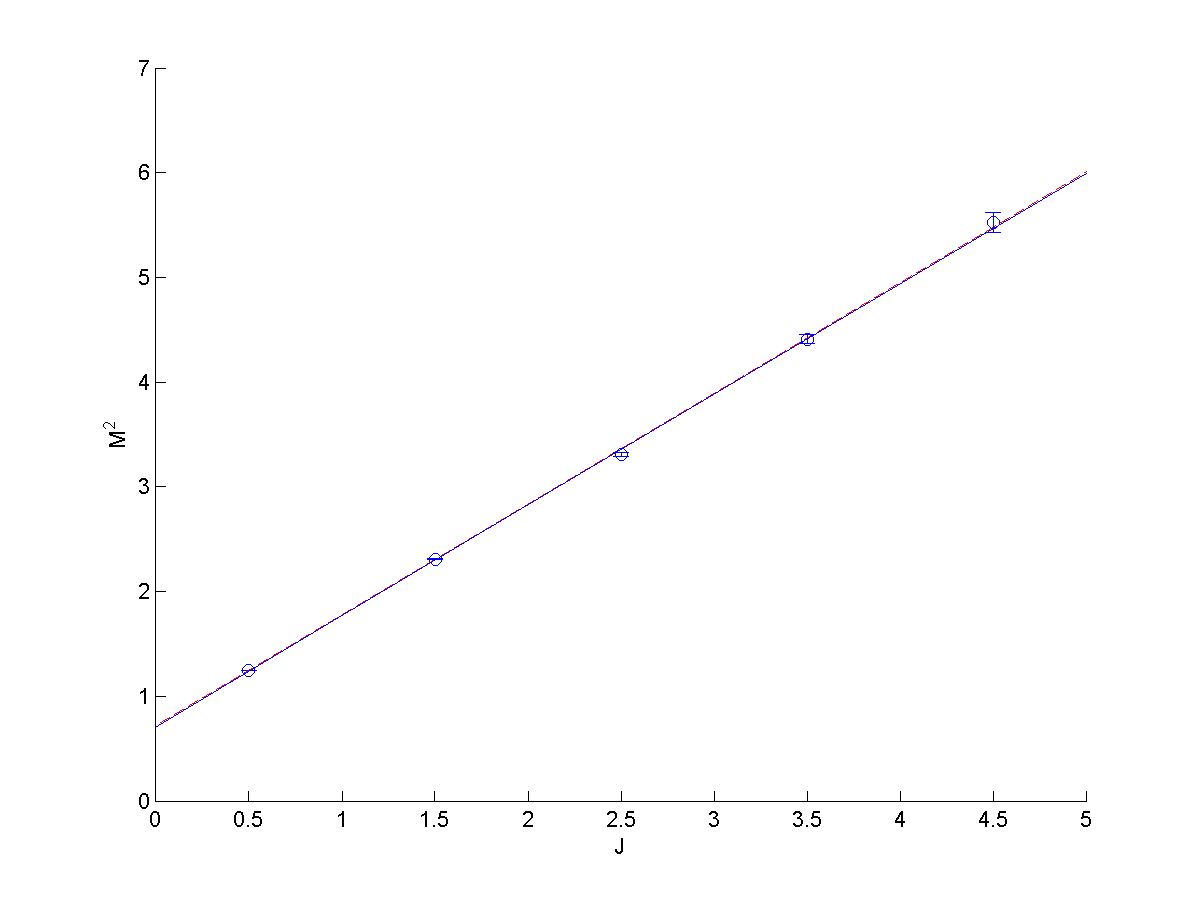}	\hfill
						\includegraphics[natwidth=1200bp, natheight=900bp, width=.48\textwidth]{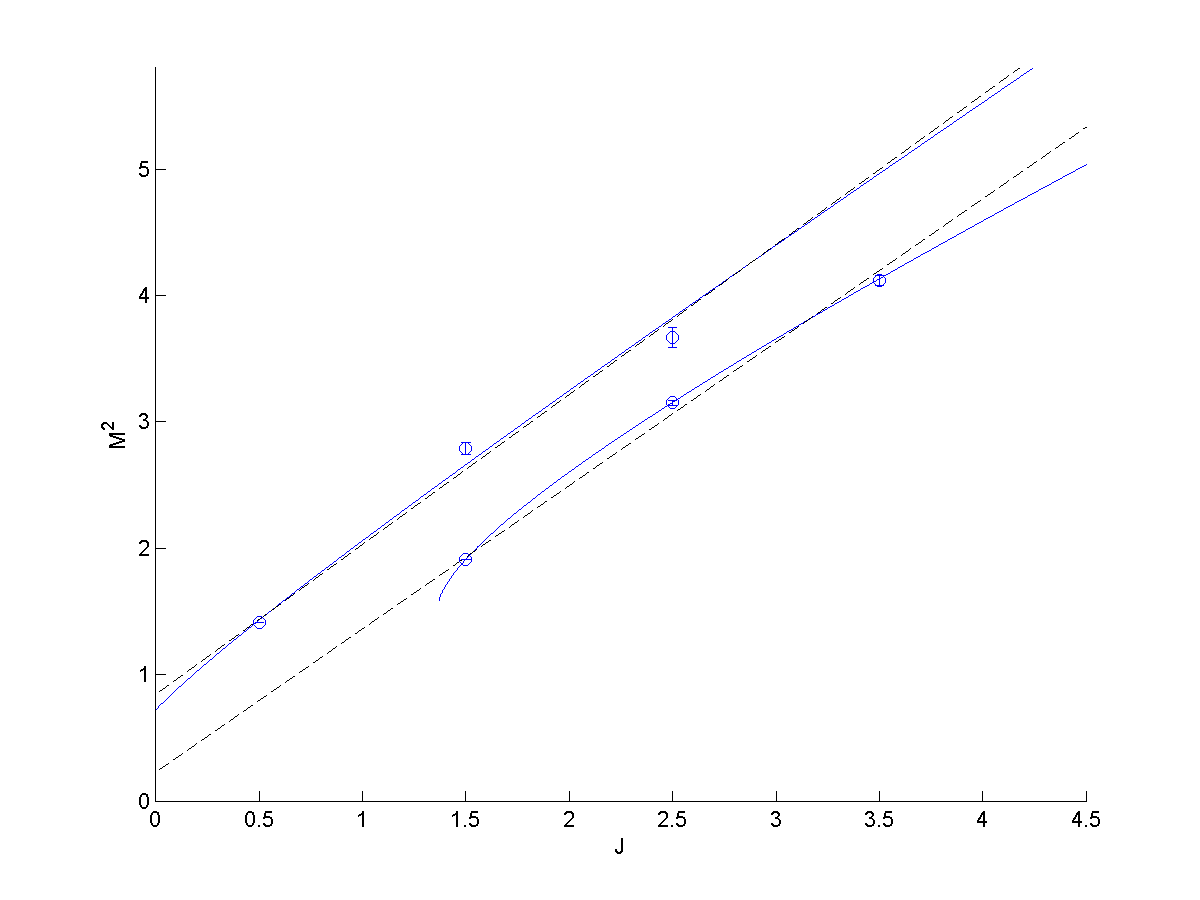} \\
						\caption{\label{fig:bar_j_strange} The strange baryon trajectories. Left: The \(\Lambda\), for which the linear fit is optimal and the maximal mass is \(2m = 125\). Right: The two \(\Sigma\) trajectories, with their massive fits - \(2m = 1190\) for the trajectory beginning with \(J^P = \jph{1}{+}\) and \(2m = 1255\) for the trajectories whose lowest state has \(J^P = \jph{3}{+}\).}
				\end{figure}
				
		\subsubsection{Strange baryons}
			\paragraph{\(I = 0\). The \(\Lambda\) trajectory:} In the left side plot of figure (\ref{fig:bar_j_strange}), we have the \(\Lambda\) baryons. These contain two light quarks (\(u/d\)), and an \(s\) quark. The states we use are \(\Lambda(1116)\jph{1}{+}\), \(\Lambda(1520)\jph{3}{-}\), \(\Lambda(1820)\jph{5}{+}\), \(\Lambda(2100)\jph{7}{-}\), and \(\Lambda(2350)\jph{9}{+}\).
			
			The even/odd effect is not present in this trajectory. The best linear fit is
			\[ \alp = 0.946, a = -0.68 \]
			with \(\chi^2_l = 0.71\ten{-4}\). Once again, it is optimal, and masses can only go up to
			\[ 2m = 125, \alp = 0.955, a = -0.61\]
			where \rchi{1.10}.
			
			\paragraph{\(I = 1\). The \(\Sigma\) trajectory:} Here we use the states \(\Sigma(1193)\jph{1}{+}\), \(\Sigma(1670)\jph{3}{-}\), and \(\Sigma(1915)\jph{5}{+}\). The best linear fit is:
			\[ \alp = 0.843, a = -0.71 \]
			with \(\chi^2_l = 26.42\ten{-4}\). The best massive fit is at the high value of
			\[ 2m = 1190, \alp = 1.502, a = 0.50 \]
			with \(\rchi{0.06}\). We also do a fit with the slope fixed at \(\alp = 0.950\), to match the result of the \(N\), \(\Delta\), and \(\Lambda\), and get as a result the fit
			\[ 2m = 620, \alp = 0.950, a = -0.15 \]
			as the optimum with \rchi{0.77}. If, instead, we assume there is an even-odd effect, then we are left only with two data points. The line connecting them is
			\[ \alp = 0.890, a_e = -0.76 \]
			and then the intercept determined from the single odd state and the given slope is \(a_o = -0.98\).
			
			\paragraph{\(I = 1\). The reverse parity \(\Sigma\) trajectory:} We have three more \(\Sigma\) states with the parity reversed relative to the states used in the previous trajectory: \(\Sigma(1385)\jph{3}{+}\), \(\Sigma(1775)\jph{5}{-}\), and \(\Sigma(2030)\jph{7}{-}\). For them, the optimal linear fit is
			\[ \alp = 0.882, a = -0.20\]
			It has \(\chi^2_l = 5.96\ten{-4}\), and the optimal massive fit,
			\[ 2m = 1255, \alp = 1.459, a = 1.37\]
			has \(\chi^2_m = 3\ten{-6}\) (\rchi{0.005}).
			The optimal fit with the slope \(\alp = 0.950\) is
			\[ 2m = 505, \alp = 0.950, a = 0.27\]
			with \rchi{0.81}. The two \(\Sigma\) trajectories are in figure (\ref{fig:bar_j_strange}).
			
\begin{figure}[t!] \centering
						\includegraphics[natwidth=1200bp, natheight=900bp, width=.48\textwidth]{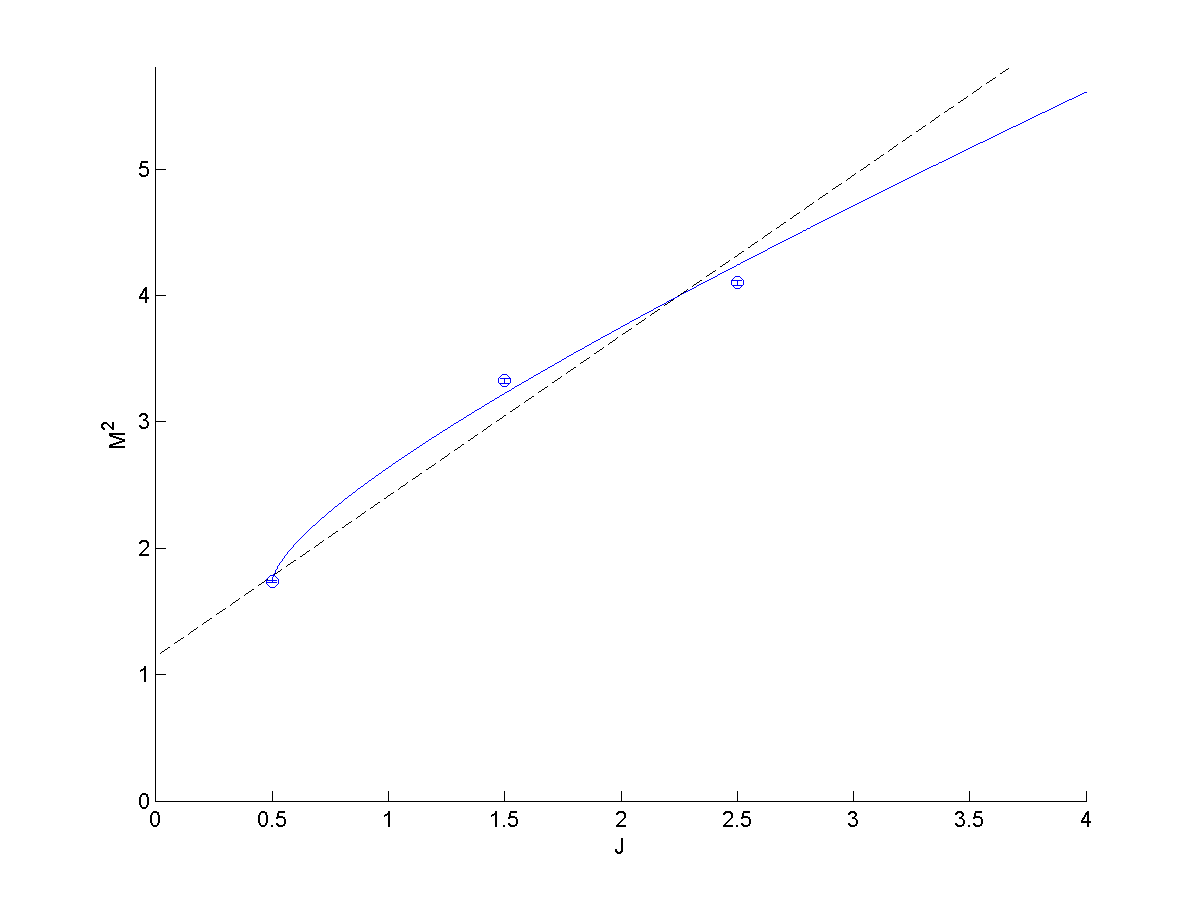}	\hfill
						\includegraphics[natwidth=1200bp, natheight=900bp, width=.48\textwidth]{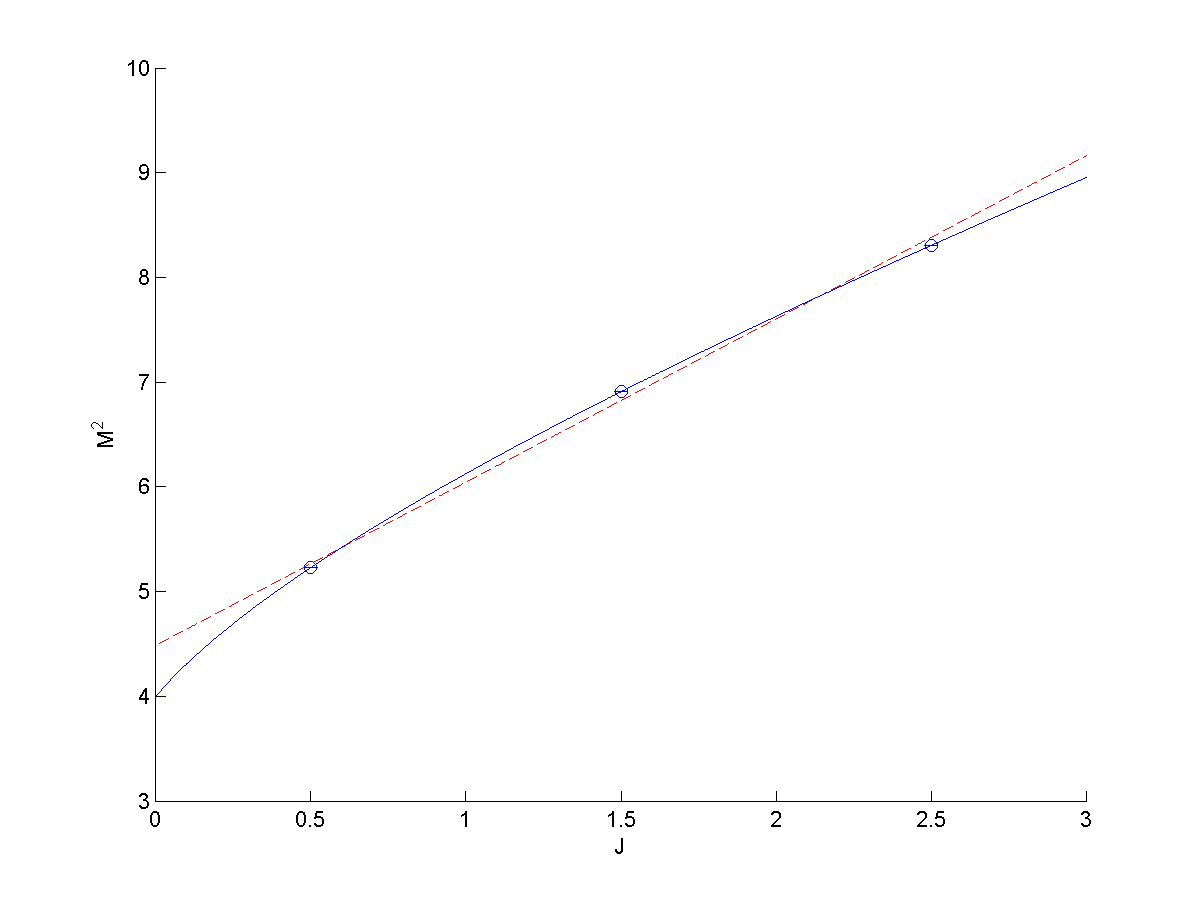} \\
						\includegraphics[natwidth=1200bp, natheight=900bp, width=.48\textwidth]{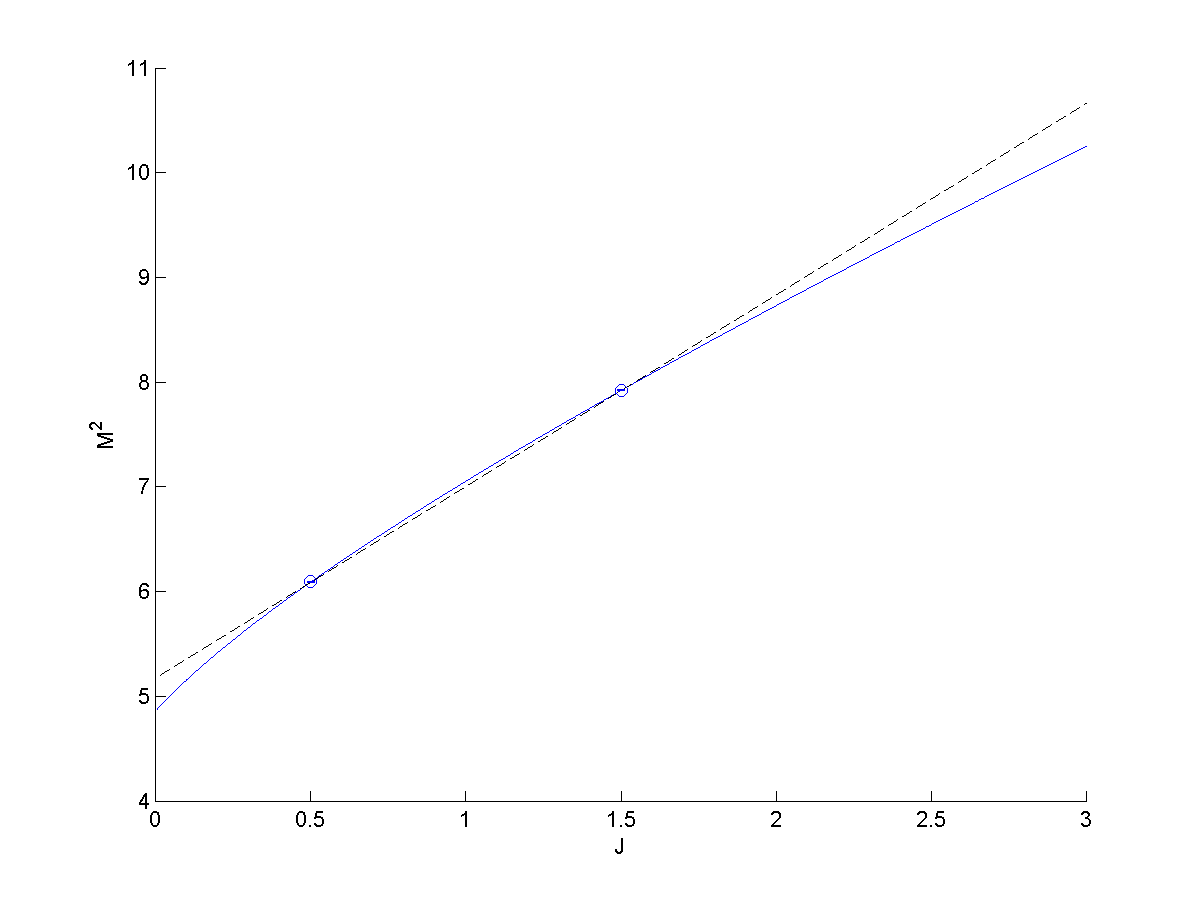}
						\caption{\label{fig:bar_j_heavy} Left: The doubly strange \(\Xi\) baryon and its fit, \(2m = 1320\). Right: The charmed \(\Lambda_c\) and its fit with \(m_1 = 90\), \(m_2 = 1720\). Bottom: The charmed-strange \(\Xi_c\) with its fit, \(2m = 2060\).}
				\end{figure}

			\paragraph{Doubly strange baryons. The \(\Xi\) trajectory:} This potential trajectory is comprised of the \(\Xi^0(1315)/\Xi^-(1322)\jph{1}{+}\), \(\Xi(1820)\jph{3}{-}\), and takes \(\Xi(2030)\) to be the \jph{5}{+} state. Its fits are plotted in figure (\ref{fig:bar_j_heavy}). The best linear fit is
			\[ \alp = 0.788, a = -0.90 \]
			with \(\chi^2_l = 51.12\ten{-4}\). The optimal symmetric fit is
			\[ 2m = 1320, \alp = 1.455, a = 0.50 \]
			with \rchi{0.20}, and the best fit with \(\alp = 0.950\) is
			\[ 2m = 850, \alp = 0.950, a = -0.04 \]
			with \rchi{0.77}.
			
		\subsubsection{Charmed baryons}
			\paragraph{The \(\Lambda_c\) trajectory:} Here we used the following three states: \(\Lambda_c(2286)^+\jph{1}{+}\), \(\Lambda_c(2625)^+\jph{3}{-}\), and \(\Lambda_c(2880)^+\jph{5}{+}\). These can be seen in the top-right plot of figure (\ref{fig:bar_j_heavy}). Assuming that, like the strange \(\Lambda\), no even-odd effect is present, the best linear fit we have is
			\[ \alp = 0.642, a = -2.88 \]
			with \(\chi^2_l = 1.39\ten{-4}\), and the optimal fit is at a high mass
			\[ 2m = 2010, \alp = 1.130, a = 0.09 \]
			with \(\chi^2_m = 2\ten{-9}\). The best fixed slope fit is
			\[ 2m = 1760, \alp = 0.950, a = -0.36 \]
			and it has \(\chi^2_m 0.24\ten{-4}\) (\rchi{0.17}). Some fits with an asymmetric distribution of the masses are
			\[ m_1 = 1720, m_2 = 90, \alp = 1.221, a = -0.13 \]
			with \(\chi^2_m = 6\ten{-10}\), or in fixed slope case, we find values near
			\[ m_1 = 1400, m_2 = 90, \alp = 0.950, a = -0.68 \]
			which has \(\chi^2_m = 0.30\ten{-4}\) (\rchi{0.22}).
			
		\subsubsection{Charmed-strange baryons: The \texorpdfstring{$\Xi_c$}{Xi-c}}
		We include here a fit done to a trajectory containing only two points. To reduce the number of fitting parameters, we do the fits with the assumption \(m_1 = m_2 = m\) and with the fixed slope \(\alp = 0.95\) GeV\(^{-2}\). This leaves us with only two fitting parameters: the total mass \(2m\) and the quantum intercept \(a\).
		The states used are the \(\Xi_c^+/\Xi_c^0\) \(\jph{1}{+}\) and \(\Xi_c(2815)\) \(\jph{3}{-}\). The best fit with the slope fixed is
		\[ 2m = 2060, \alp = 0.950, a = -0.13 \]
		The linear fit connecting the two points is
		\[ \alp = 0.547, a = -2.83 \]
		\(\chi^2 \approx 0\) for both these fits. They are plotted in figure (\ref{fig:bar_j_heavy}).

		\subsection{Trajectories in the \texorpdfstring{$(n,M^2)$}{(n,M2)} plane} \label{app:barn}
		\subsubsection{Light quark baryons}
\begin{figure}[tp] \centering
						\includegraphics[natwidth=1200bp, natheight=900bp, width=.48\textwidth]{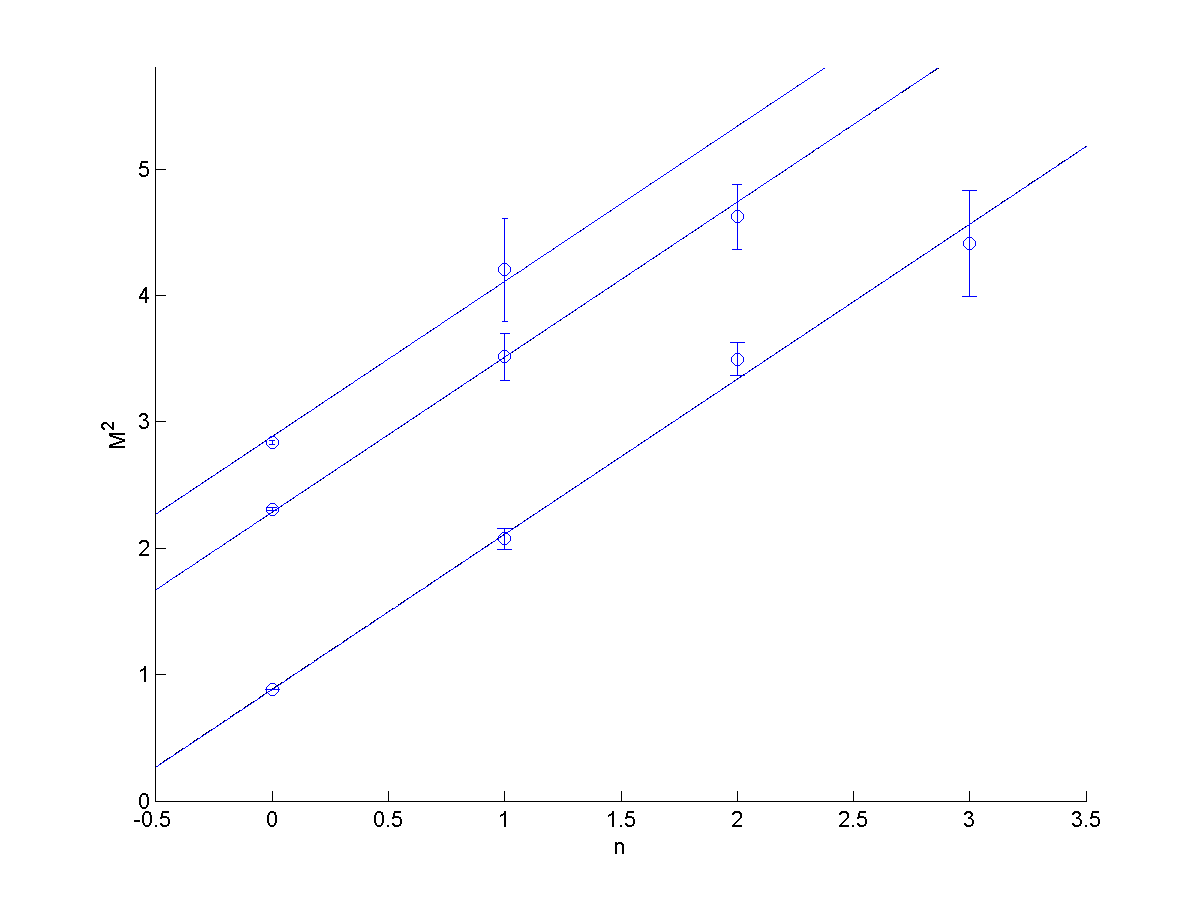}	\hfill
						\includegraphics[natwidth=1200bp, natheight=900bp, width=.48\textwidth]{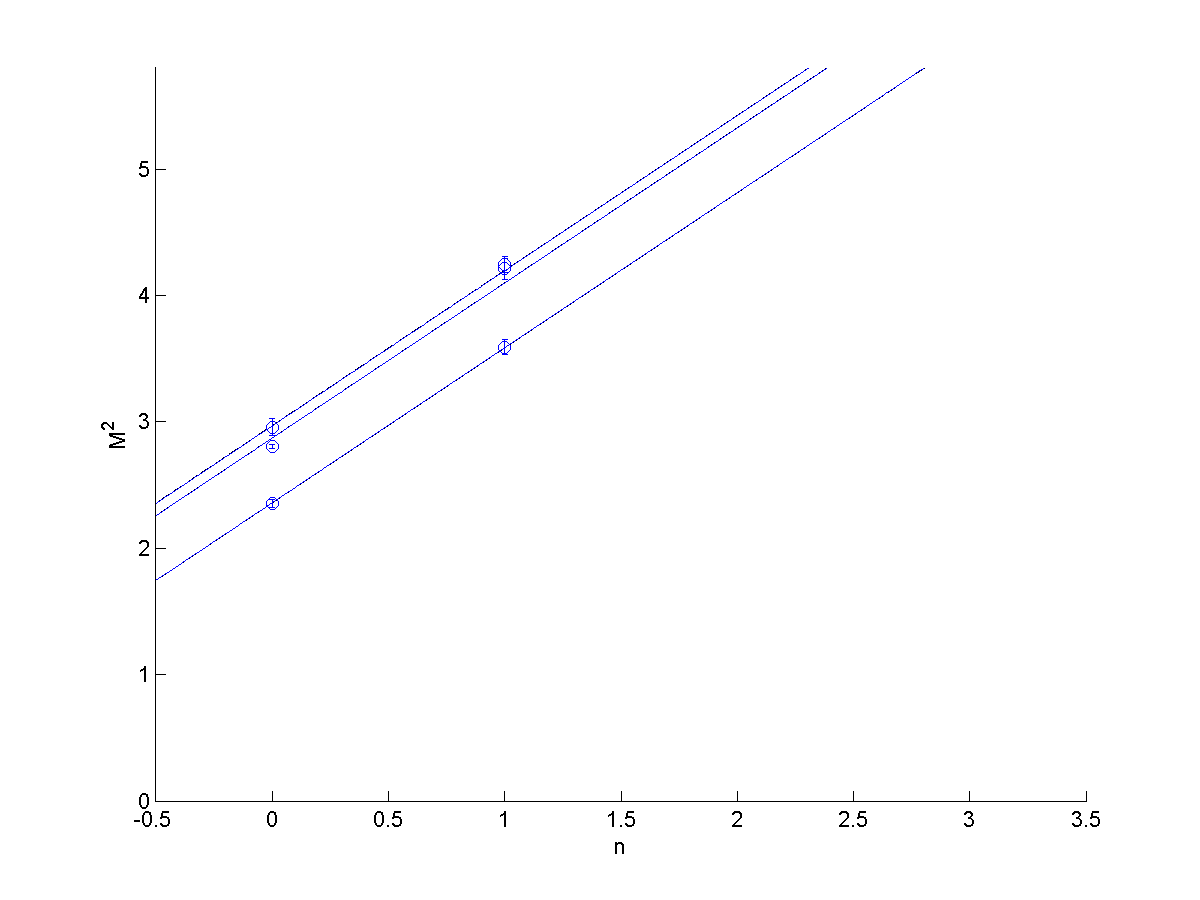} \\
						\includegraphics[natwidth=1200bp, natheight=900bp, width=.48\textwidth]{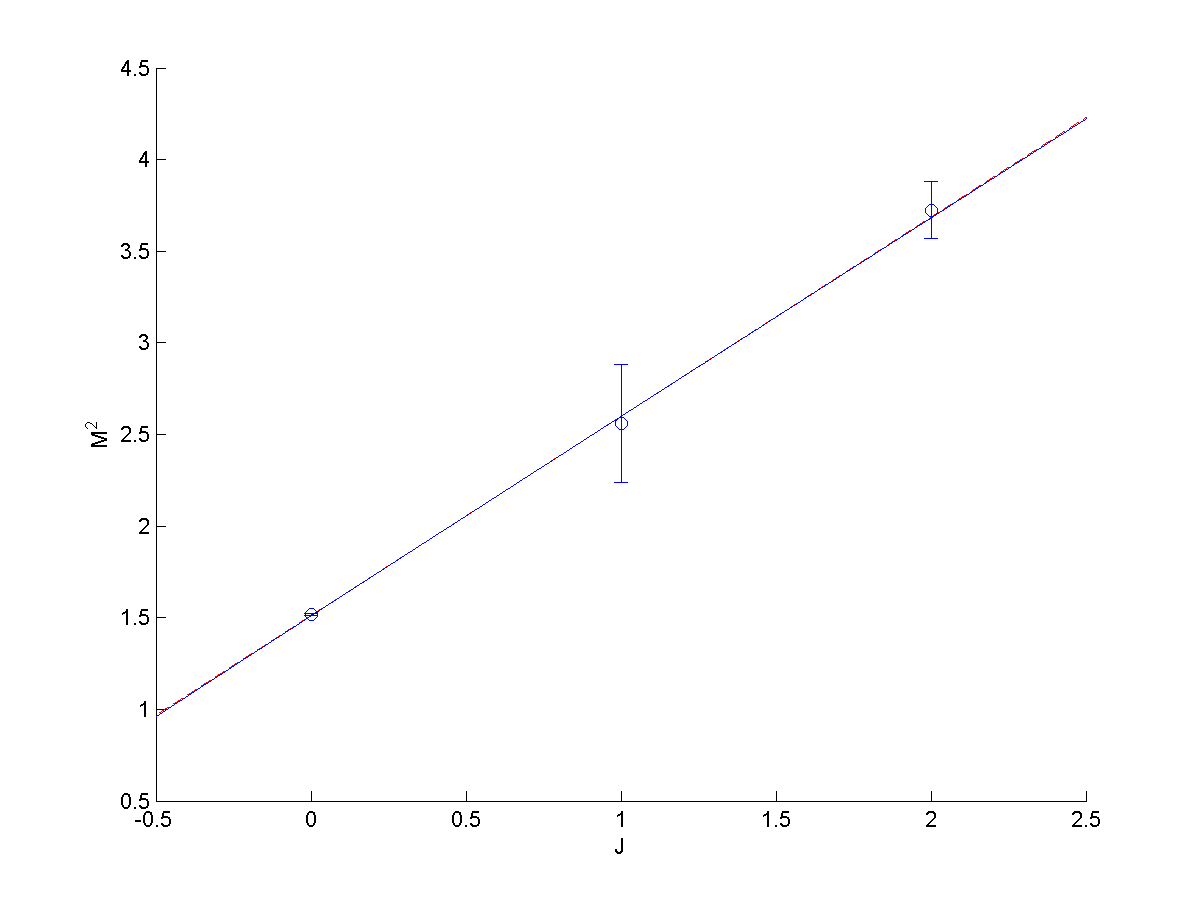}
						\caption{\label{fig:bar_n} Top: The radial trajectories of the \(N\). Top-left are the \(J^P =\) \jph{1}{+}, \jph{3}{-}, and \jph{5}{+} states. Top right are \jph{1}{-}, \jph{3}{+}, and \jph{5}{-}. All are fitted with the same slope and mass. Bottom: Radial trajectory of the \(\Delta\) baryon.}
				\end{figure}

		\paragraph{The \(N\) radial trajectory:} We use a total of 15 states belonging to six different trajectories with different \(J^P\) assignments. They are \(N(939)\), \(N(1440)\), \(N(1880)\), and \(N(2100)\) with \(J^P = \jph{1}{+}\), \(N(1520)\), \(N(1875)\), and \(N(2150)\) which have \(J^P = \jph{3}{-}\), and \(N(1680)\) and \(N(2000)\) with \(J^P = \jph{5}{+}\). Also, we have the states with reverse parity, \(N(1535)\) and \(N(1895)\) with \(\jph{1}{-}\), \(N(1720)\) and \(N(2040)\) with \(\jph{3}{+}\), and finally \(N(1675)\) and \(N(2060)\) with \(\jph{5}{-}\). The best linear fit has
		\[ \alp = 0.815\]
		\[ a_{1/2+} = -0.22, a_{3/2-} = -0.36, a_{5/2+} = 0.15\]
		\[ a_{1/2-} = -1.42, a_{3/2+} = -0.92, a_{5/2-} = 0.16 \]
		and it has \(\chi^2_l = 4.90\ten{-4}\). It is optimal and the highest good mass fit is
		\[ 2m = 425, \alp = 0.878\]
		with \(\rchi{1.10}\) and the intercepts
		\[ a_{1/2+} = 0.07, a_{3/2-} = -0.06, a_{5/2+} = 0.46\]
		\[ a_{1/2-} = -1.11, a_{3/2+} = -0.61, a_{5/2-} = 0.47 \]
		The radial trajectories are depicted in figure (\ref{fig:bar_n}).
		
		\paragraph{The \(\Delta\) radial trajectory:}	Also in figure (\ref{fig:bar_n}) we have the radial trajectory of the \(\Delta\). Here we have three states: \(\Delta(1232)\), \(\Delta(1600)\), and \(\Delta(1920)\), all with \jph{3}{+}. The linear fit
			\[ \alp = 0.920,  a = 0.11 \]
			is optimal with \(\chi^2_l = 1.78\ten{-4}\). The highest mass fits which are still close to the linear fit are around
			\[ 2m = 175, \alp = 0.936, a = 0.21\]
			with \rchi{1.10}.
			
				\clearpage

\section{Predictions for higher states}
In tables (\ref{tab:predictionsb}) and (\ref{tab:predictionsbn}) are the predictions for higher \(J\) and \(n\) baryons, based on the results of our fits.

\begin{table}[tp] \centering
					\begin{tabular}{|c|cc|cc|} \hline
						
						Trajectory & \multicolumn{4}{|c|}{Next states} \\ \hline
						
						\(N\) & \(\jph{15}{-}\): & \(2950-2960\) & \(\jph{17}{+}\): & \(3050-3060\) \\
						
						\(\Delta\) & \(\jph{17}{-}\): & \(3115-3250\) & \(\jph{19}{+}\): & \(3170-3230\) \\
						
						\(\Lambda\) & \(\jph{11}{-}\): & \(2555\) & \(\jph{13}{+}\): & \(2750-2755\) \\
						
						\(\Sigma\) & \(\jph{7}{-}\): & \(2210/2075\) & \(\jph{9}{+}\): & \(2445/2270\) \\
						
						\(\Sigma\) & \(\jph{9}{-}\): & \(2245/2295\) & \(\jph{11}{+}\): & \(2430/2520\) \\
						
						\(\Xi\) & \(\jph{7}{-}\): & \(2270/2325\) & \(\jph{9}{+}\): & \(2460/2560\) \\
						
						\(\Lambda_c\) & \(\jph{7}{-}\): & \(3095/3060\) & \(\jph{9}{+}\): & \(3280/3275\) \\
						
						\(\Xi_c\) & \(\jph{5}{+}\): & \(--/3080\) & \(\jph{7}{-}\): & \(--/3315\) \\
					\hline \end{tabular}
					\caption{\label{tab:predictionsb} Predictions for the next states in the \((J,M^2)\) plane based on the optimal massive fits, with their \(J^{PC}\) and mass (in MeV) values. The ranges listed correspond to the ranges in table (\ref{tab:summaryb}). For the \(\Sigma\) baryons, the \(\Xi\) and the \(\Lambda_c\) the first value is using the higher mass and slope of table (\ref{tab:summaryb}), and the second is the value from the \(\alp = 0.95\) fixed slope fits in table (\ref{tab:fixedslope}).}
					\end{table}
					\begin{table}[tp] \centering
					\begin{tabular}{|c|cc|cc|} \hline
						
						Trajectory & \multicolumn{4}{|c|}{Next states} \\ \hline
						
						\(N(\jp{1}{+})\) & \(n = 4\): & \(2395-2405\) & \(n = 5\): & \(2625-2650\) \\
						
						\(N(\jp{3}{-})\) & \(n = 3\): & \(2425-2440\) & \(n = 4\): & \(2655-2680\) \\
						
						\(N(\jp{5}{+})\) & \(n = 2\): & \(2295-2310\) & \(n = 3\): & \(2540-2560\) \\
						
						\(N(\jp{1}{-})\) & \(n = 2\): & \(2180-2195\) & \(n = 3\): & \(2435-2455\) \\
						
						\(N(\jp{3}{+})\) & \(n = 2\): & \(2315-2330\) & \(n = 3\): & \(2555-2580\) \\
						
						\(N(\jp{5}{-})\) & \(n = 2\): & \(2290-2310\) & \(n = 3\): & \(2535-2560\) \\
						
						\(\Delta(\jp{3}{+})\) & \(n = 3\): & \(2180-2185\) & \(n = 4\): & \(2415-2420\) \\
						
					\hline \end{tabular}
					\caption{\label{tab:predictionsbn} Predictions for the next states in the \((n,M^2)\) plane based on the optimal massive fits. Mass are in MeV. The ranges listed correspond to the ranges in table (\ref{tab:summaryb}).}
					\end{table}
					
\clearpage
\bibliographystyle{JHEP}
\bibliography{Baryons}

\providecommand{\href}[2]{#2}\begingroup\raggedright\begin{thebibliography}{10}

\bibitem{Sonnenschein:2014jwa}
J.~Sonnenschein and D.~Weissman, {\it {Rotating strings confronting PDG
  mesons}},  {\em JHEP} {\bf 1408} (2014) 013,
  [\href{http://xxx.lanl.gov/abs/1402.5603}{{\tt arXiv:1402.5603}}].

\bibitem{Kruczenski:2004me}
M.~Kruczenski, L.~A. Pando~Zayas, J.~Sonnenschein, and D.~Vaman, {\it {Regge
  trajectories for mesons in the holographic dual of large-N(c) QCD}},  {\em
  JHEP} {\bf 0506} (2005) 046,
  [\href{http://xxx.lanl.gov/abs/hep-th/0410035}{{\tt hep-th/0410035}}].

\bibitem{Inopin:1999nf}
A.~Inopin and G.~Sharov, {\it {Hadronic Regge trajectories: Problems and
  approaches}},  {\em Phys.Rev.} {\bf D63} (2001) 054023,
  [\href{http://xxx.lanl.gov/abs/hep-ph/9905499}{{\tt hep-ph/9905499}}].

\bibitem{Tang:2000tb}
A.~Tang and J.~W. Norbury, {\it {Properties of Regge trajectories}},  {\em
  Phys.Rev.} {\bf D62} (2000) 016006,
  [\href{http://xxx.lanl.gov/abs/hep-ph/0004078}{{\tt hep-ph/0004078}}].

\bibitem{Klempt:2009pi}
E.~Klempt and J.-M. Richard, {\it {Baryon spectroscopy}},  {\em Rev.Mod.Phys.}
  {\bf 82} (2010) 1095--1153, [\href{http://xxx.lanl.gov/abs/0901.2055}{{\tt
  arXiv:0901.2055}}].

\bibitem{Hata:2007mb}
H.~Hata, T.~Sakai, S.~Sugimoto, and S.~Yamato, {\it {Baryons from instantons in
  holographic QCD}},  {\em Prog.Theor.Phys.} {\bf 117} (2007) 1157,
  [\href{http://xxx.lanl.gov/abs/hep-th/0701280}{{\tt hep-th/0701280}}].

\bibitem{SakSug}
T.~Sakai and S.~Sugimoto, {\it {Low energy hadron physics in holographic QCD}},
   {\em Prog.Theor.Phys.} {\bf 113} (2005) 843--882,
  [\href{http://xxx.lanl.gov/abs/hep-th/0412141}{{\tt hep-th/0412141}}].

\bibitem{Kaplunovsky:2010eh}
V.~Kaplunovsky and J.~Sonnenschein, {\it {Searching for an Attractive Force in
  Holographic Nuclear Physics}},  {\em JHEP} {\bf 1105} (2011) 058,
  [\href{http://xxx.lanl.gov/abs/1003.2621}{{\tt arXiv:1003.2621}}].

\bibitem{Kaplunovsky:2012gb}
V.~Kaplunovsky, D.~Melnikov, and J.~Sonnenschein, {\it {Baryonic Popcorn}},
  {\em JHEP} {\bf 1211} (2012) 047,
  [\href{http://xxx.lanl.gov/abs/1201.1331}{{\tt arXiv:1201.1331}}].

\bibitem{Kaplunovsky:2013iza}
V.~Kaplunovsky and J.~Sonnenschein, {\it {Dimension Changing Phase Transitions
  in Instanton Crystals}},  {\em JHEP} {\bf 1404} (2014) 022,
  [\href{http://xxx.lanl.gov/abs/1304.7540}{{\tt arXiv:1304.7540}}].

\bibitem{Witten:1998zw}
E.~Witten, {\it {Anti-de Sitter space, thermal phase transition, and
  confinement in gauge theories}},  {\em Adv.Theor.Math.Phys.} {\bf 2} (1998)
  505--532, [\href{http://xxx.lanl.gov/abs/hep-th/9803131}{{\tt
  hep-th/9803131}}].

\bibitem{Brandhuber:1998xy}
A.~Brandhuber, N.~Itzhaki, J.~Sonnenschein, and S.~Yankielowicz, {\it {Baryons
  from supergravity}},  {\em JHEP} {\bf 9807} (1998) 020,
  [\href{http://xxx.lanl.gov/abs/hep-th/9806158}{{\tt hep-th/9806158}}].

\bibitem{Callan:1999zf}
J.~Callan, Curtis~G., A.~Guijosa, K.~G. Savvidy, and O.~Tafjord, {\it {Baryons
  and flux tubes in confining gauge theories from brane actions}},  {\em
  Nucl.Phys.} {\bf B555} (1999) 183--200,
  [\href{http://xxx.lanl.gov/abs/hep-th/9902197}{{\tt hep-th/9902197}}].

\bibitem{Dymarsky:2009cm}
A.~Dymarsky, S.~Kuperstein, and J.~Sonnenschein, {\it {Chiral Symmetry Breaking
  with non-SUSY D7-branes in ISD backgrounds}},  {\em JHEP} {\bf 0908} (2009)
  005, [\href{http://xxx.lanl.gov/abs/0904.0988}{{\tt arXiv:0904.0988}}].

\bibitem{Seki:2008mu}
S.~Seki and J.~Sonnenschein, {\it {Comments on Baryons in Holographic QCD}},
  {\em JHEP} {\bf 0901} (2009) 053,
  [\href{http://xxx.lanl.gov/abs/0810.1633}{{\tt arXiv:0810.1633}}].

\bibitem{Hellerman:2013kba}
S.~Hellerman and I.~Swanson, {\it {String Theory of the Regge Intercept}},
  \href{http://xxx.lanl.gov/abs/1312.0999}{{\tt arXiv:1312.0999}}.

\bibitem{Chodos:1973gt}
A.~Chodos and C.~B. Thorn, {\it {Making the Massless String Massive}},  {\em
  Nucl.Phys.} {\bf B72} (1974) 509.

\bibitem{Baker:2002km}
M.~Baker and R.~Steinke, {\it {Semiclassical quantization of effective string
  theory and Regge trajectories}},  {\em Phys.Rev.} {\bf D65} (2002) 094042,
  [\href{http://xxx.lanl.gov/abs/hep-th/0201169}{{\tt hep-th/0201169}}].

\bibitem{Zahn:2013yma}
J.~Zahn, {\it {The excitation spectrum of rotating strings with masses at the
  ends}},  {\em JHEP} {\bf 1312} (2013) 047,
  [\href{http://xxx.lanl.gov/abs/1310.0253}{{\tt arXiv:1310.0253}}].

\bibitem{Witten:1998xy}
E.~Witten, {\it {Baryons and branes in anti-de Sitter space}},  {\em JHEP} {\bf
  9807} (1998) 006, [\href{http://xxx.lanl.gov/abs/hep-th/9805112}{{\tt
  hep-th/9805112}}].

\bibitem{Dymarsky:2010ci}
A.~Dymarsky, D.~Melnikov, and J.~Sonnenschein, {\it {Attractive Holographic
  Baryons}},  {\em JHEP} {\bf 1106} (2011) 145,
  [\href{http://xxx.lanl.gov/abs/1012.1616}{{\tt arXiv:1012.1616}}].

\bibitem{Kuperstein:2004yf}
S.~Kuperstein and J.~Sonnenschein, {\it {Non-critical, near extremal AdS(6)
  background as a holographic laboratory of four dimensional YM theory}},  {\em
  JHEP} {\bf 0411} (2004) 026,
  [\href{http://xxx.lanl.gov/abs/hep-th/0411009}{{\tt hep-th/0411009}}].

\bibitem{'tHooft:2004he}
G.~'t~Hooft, {\it {Minimal strings for baryons}},
  \href{http://xxx.lanl.gov/abs/hep-th/0408148}{{\tt hep-th/0408148}}.

\bibitem{Sharov:2000pg}
G.~Sharov, {\it {Quasirotational motions and stability problem in dynamics of
  string hadron models}},  {\em Phys.Rev.} {\bf D62} (2000) 094015,
  [\href{http://xxx.lanl.gov/abs/hep-ph/0004003}{{\tt hep-ph/0004003}}].

\bibitem{ThesisFederovsky}
E.~Federovsky, {\it {Stringy baryons and scattering amplitudes}},  Master's
  thesis, Tel Aviv University, August, 2010.

\bibitem{ThesisHarpaz}
G.~Harpaz, {\it {Simulating stringy baryons}},  Master's thesis, Tel Aviv
  University, June, 2008.

\bibitem{Sharov:2013tga}
G.~Sharov, {\it {String Models, Stability and Regge Trajectories for Hadron
  States}},  \href{http://xxx.lanl.gov/abs/1305.3985}{{\tt arXiv:1305.3985}}.

\bibitem{Selem:2006nd}
A.~Selem and F.~Wilczek, {\it {Hadron systematics and emergent diquarks}},
  \href{http://xxx.lanl.gov/abs/hep-ph/0602128}{{\tt hep-ph/0602128}}.

\bibitem{Friedmann:2014qpa}
T.~Friedmann, {\it {QCD vs. the Centrifugal Barrier: a New QCD Effect}},  {\em
  EPJ Web Conf.} {\bf 70} (2014) 00027.

\bibitem{PDG:2012}
{\bf Particle Data Group} Collaboration, J.~Beringer et~al., {\it {Review of
  Particle Physics (RPP)}},  {\em Phys.Rev.} {\bf D86} (2012) 010001.

\bibitem{Sharov:1998hi}
G.~Sharov, {\it {String baryon model 'triangle': Hypocycloidal solutions}},
  {\em Phys.Rev.} {\bf D58} (1998) 114009,
  [\href{http://xxx.lanl.gov/abs/hep-th/9808099}{{\tt hep-th/9808099}}].

\bibitem{Peeters:2005fq}
K.~Peeters, J.~Sonnenschein, and M.~Zamaklar, {\it {Holographic decays of
  large-spin mesons}},  {\em JHEP} {\bf 0602} (2006) 009,
  [\href{http://xxx.lanl.gov/abs/hep-th/0511044}{{\tt hep-th/0511044}}].

\bibitem{Decays}
J.~Sonnenschein and D.~Weissman, {\it {On decays of stringy hadrons [work in
  progress]}}, .

\end{thebibliography}\endgroup
\end{document}